# Chain-length-dependent correlated molecular motion in polymers

Matthew Reynolds[1], Daniel L. Baker[1], Peter D. Olmsted[2], and Johan Mattsson[1*]
[1]*School of Physics and Astronomy, University of Leeds, Leeds LS2 9JT, United Kingdom and*
[2]*Department of Physics and Institute for Soft Matter Synthesis and Metrology, Georgetown University, Washington DC, 20057*

We show how dynamic heterogeneities (DH), a hallmark of glass-forming materials, depend on chain flexibility and chain length in polymers. For highly flexible polymers, a relatively large number of monomers ($N_c \sim 500$) undergo correlated motion at the glass transition temperature ($T_g$), independent of molecular weight ($M$). In contrast, less flexible polymers show a complex $N_c(M)$ behaviour divided into three regimes, consistent with observations in both $T_g(M)$ and chain conformational structure. For short oligomers ($\lesssim 2$ Kuhn steps), a transition from mainly *inter*molecular correlations and $N_c \sim 200$, to strongly *intra*molecular correlations and $N_c < 50$ (roughly the molecular size) is observed; for longer chains, $N_c$ increases weakly, before saturating. For poly(methyl methacrylate), a remarkable similarity is found between $N_c(M)$ and the $M$-dependent ratio of the activation barriers of the structural ($\alpha$) and secondary ($\beta$) relaxations; we present evidence that this relationship is a general feature of glass-forming polymers. Our results suggest a direct link between the DH length-scale and the number of $\beta$ relaxation events jointly-activated to facilitate the $\alpha$ relaxation.

*Introduction*— Molecular motions in polymer melts slow down during cooling, eventually resulting in the formation of a disordered solid – a glass. The glass transition temperature $T_g$ is defined as $\tau_\alpha(T_g) \equiv 100$s [1], where $\tau_\alpha(T)$ is the temperature ($T$) dependent structural ($\alpha$) relaxation time [2]. Near $T_g$, so called dynamic heterogeneities (DH), in which the local dynamics vary from place to place, become more prominent [3, 4]. Their characteristic length-scale $\xi_{\rm DH}$ has been determined for both polymeric and non-polymeric liquids [5–9], using either experimental [9–16] or computational [17–23] approaches, with a resulting length-scale $\xi_{\rm DH}(T_g) \sim$ 1-5 nm [24], which for polymers correspond to 50-500 monomers [7, 9, 25]. However, very few studies have investigated how polymer chain-length affects DH [9, 26, 27] and the entire range from short oligomers to long-chain polymers has rarely been explored [27].

The glass transition of polymers and its associated dynamics can typically be divided into three distinct regimes [28, 29], corresponding to (I) short oligomers ($\lesssim 2$ Kuhn steps), (II) chains with $\sim$ 2-20 Kuhn steps, and (III) long chains. These regimes are captured in the $T_g(M)$ behaviour (segmental dynamics) and in the chain structure [29]. The inset to Fig. 1 shows an example for PMMA, illustrating $T_g(M)$ (open black symbols) and the $M$-dependent single-molecule aspect ratio $\Lambda^2(M) = \lambda_3(M)^2/\lambda_1(M)^2$ (solid green symbols), where $\lambda_3^2$ and $\lambda_1^2$ are the largest and smallest eigenvalues, respectively, of the average conformational tensor [29]. The regime I-II boundary at $M^\star$ is clearly manifested as both a kink in $T_g(M)$ and a peak in $\Lambda^2$, where the latter is due to chain folding occurring for $M \gtrsim M^\star$ [29]. For long chains in regime III ($> M^{\star\star}$), $\Lambda^2$ approaches the Gaussian chain value of $\approx$12.

To determine how chain-length (or molecular weight $M$) affects DH in polymers, we study four different polymer chemistries: poly(methyl methacrylate) (PMMA); polystyrene (PS); poly(propylene glycol dimethyl ether) (PPG-DME), and poly(dimethyl siloxane) (PDMS). PMMA and PS show relatively high $T_g$, dynamic fragility $m$ (the sensitivity of $\tau_\alpha(T)$ to a $T$-variation near $T_g$), and chain stiffness [30], whereas PPG-DME and PDMS are significantly more flexible polymers characterised by lower $T_g$ and $m$.

DH can be characterised by a "four-point" correlation function $G_4(\boldsymbol{r}, t)$ that correlates the relaxation dynamics (probed by two-point correlators) in space and time. A space-integration of $G_4(\boldsymbol{r}, t)$ yields a dynamic susceptibility $\chi_4(t)$, which quantifies the fluctuations around the average dynamics [4, 19, 22, 31, 32]. $\chi_4(t)$ is typically a non-monotonic function with a peak $\chi_4^{\rm max}$ near $t_{\rm max} \simeq \tau_\alpha(T)$, where $\chi_4^{\rm max}$ is proportional to the volume of correlated motions, or correspondingly the number of monomer units $N_c^{(4)}$ that undergo correlated motions.

Direct determination of $\chi_4(t)$ is difficult and has mainly been achieved in simulations [4, 19, 22] and for colloidal systems [33, 34], and recently for a metallic glass-former [35]. However, it has been demonstrated that $\chi_4(t)$ can be estimated from the temperature dependence of a dynamic susceptibility [31, 32], obtained from broadband dielectric spectroscopy (BDS), rheology, or scattering [31, 32].

We employ two approaches to determine the number of dynamically correlated units involved in the $\alpha$ relaxation (often referred to as a cooperatively rearranging region, or a CRR). The approach of Donth [11, 36] yields the number of correlated units $N_c(T \sim T_g)$ from the thermal fluctuations measured near $T_g$ using temperature modulated differential scanning calorimetry (TMDSC). The approach of Berthier *et al.* [31] estimates $\chi_4$ from the temperature-dependence of the complex permittiv-ity measured using BDS, which then yields $N_c(T)$. We use both methods to determine the $M$-dependence of $N_c$ for the four investigated polymers, and show in the End Matter that both approaches give consistent quantitative results.

* k.j.l.mattsson@leeds.ac.uk

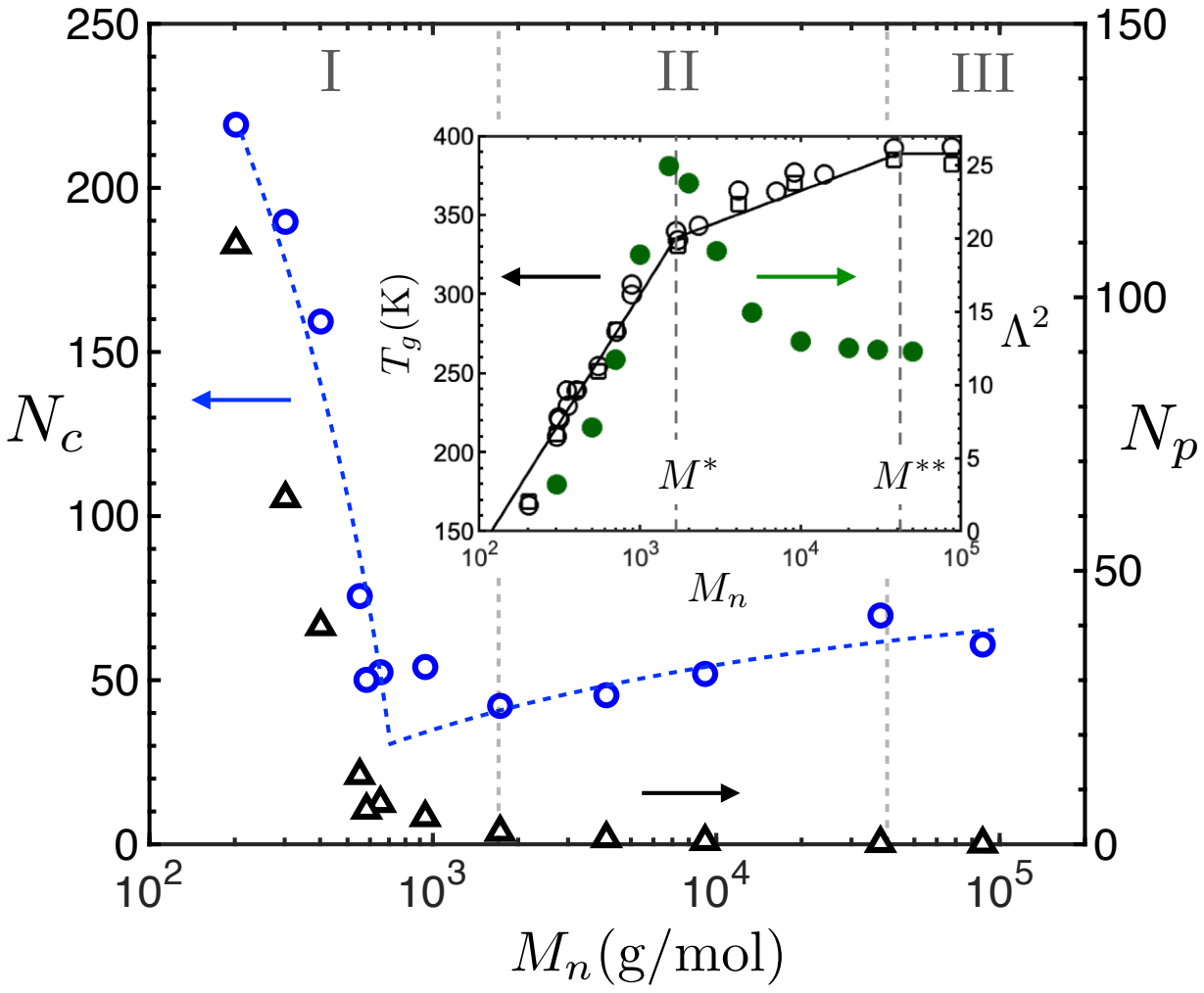

FIG. 1. (Left; blue circles) The number of monomers $N_c$ involved in the $\alpha$ relaxation at $T_\alpha \simeq T_g$ for PMMA, determined using TMDSC. (Right; black triangles) $N_p \equiv N_c/n$, where $n$ is the degree of polymerisation. For short chains, where the behaviour is intermolecular, $N_p$ is roughly the number of molecules involved in the $\alpha$ relaxation. In contrast, $N_p \sim 1$ for $M \sim M^*$, which demonstrates the importance of intramolecular dynamics for $M \geq M^*$. The dashed (blue) lines are guides to the eye. The dotted vertical lines mark the regime boundaries at $M^*$ and $M^{**}$, as shown in the inset. The inset shows the $M$-dependence of $T_g$ (left) and the average polymer aspect ratio $\Lambda^2$ (right), from Baker *et al.* [29]. The dotted vertical lines in the main panel and the inset mark the regime boundaries at $M^*$ and $M^{**}$.

*Correlated Dynamics*— The results for $N_c(M)$ (at $T_\alpha \simeq T_g$; see End Matter for a detailed discussion) are shown for PMMA in blue rings in Figure 1. We find that $N_c(M)$ falls dramatically in regime I from $N_c \sim 220$ for the dimer to $N_c < 50$ as the regime II boundary at $M^\star$ is approached. This demonstrates that near $T_g$ the $\alpha$ relaxation dynamics for short oligomers (within regime I) are highly correlated, and since these molecules are very short, the dynamics are strongly intermolecular. This is further illustrated in black triangles in Fig. 1, which shows $N_\text{p}(M) = N_c(M)/n(M)$, where $n$ is the degree of polymerisation. For short chains $N_p(M)$ is roughly the $M$-dependent number of correlated molecules. We observe a drop from $N_p \sim 110$ for the dimer to $N_p \sim \mathcal{O}(1)$ near $M^\star$, reflecting strongly intramolecular behaviour for $M \gtrsim M^\star$. Figure 1 shows that $N_c$ increases slightly with $M$ in regime II to reach $N_c \sim 60$ in regime III. Thus, the number of correlated monomers $N_c(M)$, as well as $T_g(M)$ and the chain structure $\Lambda^2(M)$ (inset of Fig. 1) all show distinct changes in behaviour for similar characteristic molecular weights.

We compare $N_c(M)$ for all four polymers in Fig. 2. The stiffer polymers (PMMA and PS) show a sharp decrease of $N_c(M)$ in regime I, followed by a weak increase in regime II. However, the more flexible polymers (PPG-DME and PDMS) show a roughly constant $N_c(M)$, which exceeds that of the shortest PMMA and PS oligomers. For PPG-DME and PDMS, $N_c$ is larger than the degree of polymerisation $n$, *i.e.* $N_\text{p} > 1$, until regime III is reached.

These results suggest that intramolecular constraints are much weaker for PDMS and PPG-DME, so that chain connectivity and chain-length do not significantly influence their $\alpha$ relaxation behaviour, as demonstrated by the nearly $M$-independent $N_c \sim$ 300-700 $\gg n$. Consistent with this, the rotational dihedral barriers of PPG-DME and PDMS (respectively $\lesssim 1.0$ [37] and 0.6 kcal/mole [38]) are significantly smaller than those of PMMA and PS (respectively 2.8 [39] and 3 kcal/mole [38]), and the O-Si-O bending energy for PDMS is notoriously weak [40].

For non-polymeric glass-formers and for short oligomers (in regime I), the molecular motions linked to the structural $\alpha$ relaxation are dominated by translational degrees of freedom (DOF). As chains with relatively high dihedral barriers (such as PMMA or PS) grow, the number of DOF available for molecular motions is reduced, by exchanging three intermolecular translational DOF for one dihedral intramolecular DOF (per additional degree of polymerization) [41]. This is accompanied by a change of the $\alpha$ relaxation character from mainly intermolecular to highly intramolecular. This is directly illustrated near the regime I-II crossover ($M \sim M^\star$), where chain folding takes place [29] and $N_c \sim n$. Correspondingly, as chains grow within regime I, less cooling is required for dynamic arrest to occur, meaning that fewer monomers are involved in correlated motions at $T_g$, as shown in Fig 2.

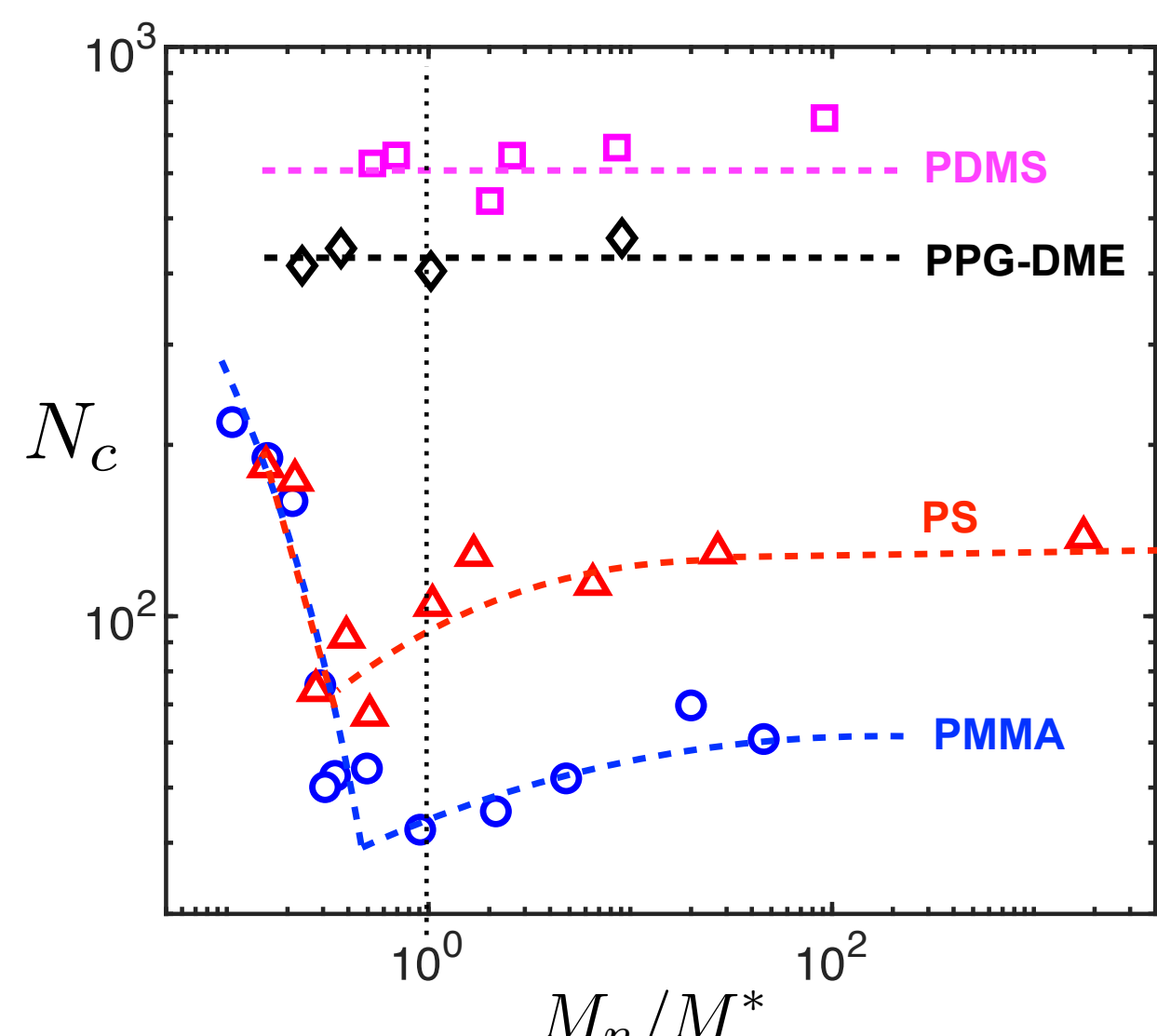

FIG. 2. Comparison of the number of correlated monomers $N_c$ involved in the $\alpha$ relaxation for four different polymers as a function of molecular weight $M_n$, scaled by the molecular weight $M^*$ separating regimes I and II.

*Activation Barriers—* To further investigate the $M$-dependence of $N_c$ observed for PMMA and PS (Fig. 2), we first note that in addition to the structural $\alpha$ relaxation, glass-formers generally also show a faster $\beta$ relaxation corresponding to more 'local' cooperative motions [29]; there is significant evidence that the two relaxations are intrinsically linked [29, 42].

Molecular relaxations in the glassy state typically follow the Arrhenius law, $\tau_i = \tau_{0i} \exp[\Delta H_i/RT]$ ($i = \alpha, \beta$), where $\tau_{0i} \sim 0.1\,\mathrm{ps}$ is a microscopic relaxation time and $R$ is the gas constant. Since $\tau_\alpha(T_g) \equiv 100\,\mathrm{s}$, $\Delta H_\alpha\ (M)|_{T=T_g} = RT_g \ln(100\mathrm{s}/\tau_0)$; while $\Delta H_\beta$ is $T$-independent. Thus, $\Delta H_\alpha(M)|_{T=T_g}$ is proportional to $T_g(M)$, and both show three regimes in $M$ [29] (inset to Fig. 1). While $\Delta H_\alpha(M)$ follows the $M$-dependence of $T_g(M)$, $\Delta H_\beta(M)$ for PMMA increases with $M$ in regime I and is nearly $M$-independent in regimes II and III [29]. Hence, for PMMA the ratio $\mathcal{R}(M)$ obeys the $M$-dependencies of both barriers in regime I, but mainly that of the $\alpha$ barrier in regimes II and III. Remarkably, as shown in Fig. 3, the ratio $\mathcal{R}(\mathcal{M}) \equiv \Delta H_\alpha(M)/\Delta H_\beta(M)$ between the two activation barriers has an $M$-dependence very similar to that of $N_c(M)$. Notably, $\mathcal{R}(M) \approx 1$ for $M \sim M^\star$, which corresponds to the onset of strongly intramolecular behaviour in $N_c$ (Fig. 1), and suggests a close correspondence between the properties of the $\alpha$ and the more 'local' $\beta$ relaxation (see a detailed discussion in [29]).

As discussed in the SI (Section E [43]), the similarities between the $M$-dependence of $N_c(M)$ and the barrier ratio $\mathcal{R}(M)$ extend beyond PMMA. By combining our own data with the small amount of literature data, we find that $R(M)$ for PS and poly($\alpha$ methyl styrene) behave similarly to $R(M)$ for PMMA. Moreover, for the highly flexible polymers PDMS and PPG-DME, distinct $\beta$ relaxations are not observed, but these polymers instead show excess wings near $T_g$ indicating highly 'merged' $\beta$ and $\alpha$ relaxations [44, 45]. As discussed in [43], we expect that such behavior should result in an effectively $M$-independent $\mathcal{R}(M)$, which is consistent with the $M$-independent behavior of $N_c(M)$ for PPG-DME and PDMS (Fig. 2). Hence, the observed correlation between $\mathcal{R}(M)$ and $N_c(M)$ appears to be general for polymers.

Both non-polymeric and long-chain (regime III) polymeric glass-formers often satisfy $\Delta H_\beta \approx 24RT_g$ [29, 42, 46], which suggests a typical value of the ratio, $\mathcal{R} \simeq \frac{1}{24}\ln(100\,\mathrm{s}/\tau_0) \approx 1.4$, and a direct link between the $\alpha$ and $\beta$ relaxations. In fact, as long ago as 1940, Kauzmann and Eyring [47] suggested that flow in polymers results from "flow segments" comprising ∼5-10 bonds. Numerous studies have since tried to link 'local' dynamics, typically on the scale of the flow segment, to the structural $\alpha$ relaxation [48–53].

*Dynamic Facilitation—* We recently suggested that the link between the $\alpha$ and $\beta$ relaxations could be explained by *dynamic facilitation* (DF) [23, 54, 55], whereby a 'local' relaxation facilitates adjacent relaxations [23, 29]. In polymers near $T_g$, we can apply DF to two situations: (1) For oligomers of stiffer polymers (such as PMMA or PS) at $T \sim T_g$, dense packing and relatively large dihedral barriers restrict 'local' intramolecular motion to arise from cooperative motion, involving a few adjacent dihedral angles. These 'local' cooperative rearrangements can propagate along the chain (*intramolecular facilitation*) and facilitate the rearrangement of the entire oligomer, resulting in the $\beta$ relaxation and its associated enthalpy $\Delta H_\beta(M)$. (2) This newly-mobile oligomeric chain can facilitate the mobility of other oligomers through *intermolecular facilitation*, to yield the $\alpha$ relaxation and its associated activation enthalpy $\Delta H_\alpha(T_g)$. For longer chains, chain folding divides the chain into $\beta$ relaxation "beads", leading to a near constant $\Delta H_\beta(M^\star)$ for $M \geq M^\star$. Here, the structural $\alpha$ relaxation (and thus $T_g$) results from propagation of mobility through either intramolecular (along the chain) or intermolecular facilitation of the $\beta$ beads; the nature of the facilitated dynamic coupling varies with chain-length, thus separating $T_g(M)$ into distinct dynamic regimes (see schematic illustration in Fig. 3).

A hallmark of DF is hierarchical relaxations [54, 55], where relaxation on one length-scale facilitates adjacent relaxation on a larger length-scale, leading to a logarithmic relationship between the length-scale $\ell$ separating fundamental excitations (relaxations) with energy $\Delta E_\sigma$, and the activation barrier $\Delta E = \Delta E_\sigma[1 + \nu \log(\ell/\sigma)]$ of the resulting facilitated relaxation, where $\nu \sim \mathcal{O}(1)$ is a constant, $\ell(M)$ is the average length-scale between fundamental relaxations, and $\sigma(M)$ is their size [23]. By applying this reasoning to $\Delta H$ data from BDS [29], we recently suggested that the ratio of $\alpha$ and $\beta$ barriers at $T_g$ obeys

$$\mathcal{R}(M) \equiv \frac{\Delta H_\alpha(M)}{\Delta H_\beta(M)} = \left[1 + \nu \log\left(\frac{\ell\,(M)}{\sigma\,(M)}\right)\right], \quad (1)$$

which is proportional to the maximum number of $\beta$ events that act in sync to facilitate the $\alpha$ relaxation [29, 55]. This is in the spirit of Fujimori and Oguni [56]'s suggestion that $\mathcal{R}$ varies with the size of the correlation region in glass formers. The simple long-chain relation $\mathcal{R} \simeq \frac{1}{24}\log(100\,\mathrm{s}/\tau_0)$ breaks down at smaller molecular weights ($M < M^{\star\star}$) in stiff polymers, where intramolecular facilitiation plays a role in the structural relaxation.

*Discussion—* The remarkable similarity between the $M$-dependence of the barrier ratio $\mathcal{R}(M)$ and $N_c(M)$ is more difficult to interpret, and even though dynamic facilitation (DF) directly implies dynamic heterogeneities (DH) [23], DH can exist without DF [19]. However, the growing evidence for DF in glass-forming systems [22, 23, 29, 57–60] makes it increasingly important to identify links between DF-properties, such as $\mathcal{R}(M)$, and the size of DH, as reflected in $N_c(M)$ or the corresponding characteristic length-scale $\xi_{\mathrm{DH}}$ [61]. Computer simulations and experiments [22, 62–64] have investigated the $T$-dependent geometry of DH, with several studies finding that dynamic clusters (CRRs) become increasingly compact near $T_g$, so that the characteristic length-scale $\xi_{\mathrm{DH}} \propto N_c^\gamma$ with $\gamma \approx 1/3$. If we use the simple approximation of compact CRRs, then the corresponding length $\xi_{\mathrm{DH}}(M) \sim N_c(M)^{-1/3}$ roughly scales with $\mathcal{R}(M)$ (Fig. 3; See Fig. S8 of [43] for more discussion). Following the DF interpretation [29, 55], the ratio $\mathcal{R}$ corresponds to the maximum number of $\beta$ relaxation 'beads' that need to be jointly activated to facilitate the structural $\alpha$ relaxation.

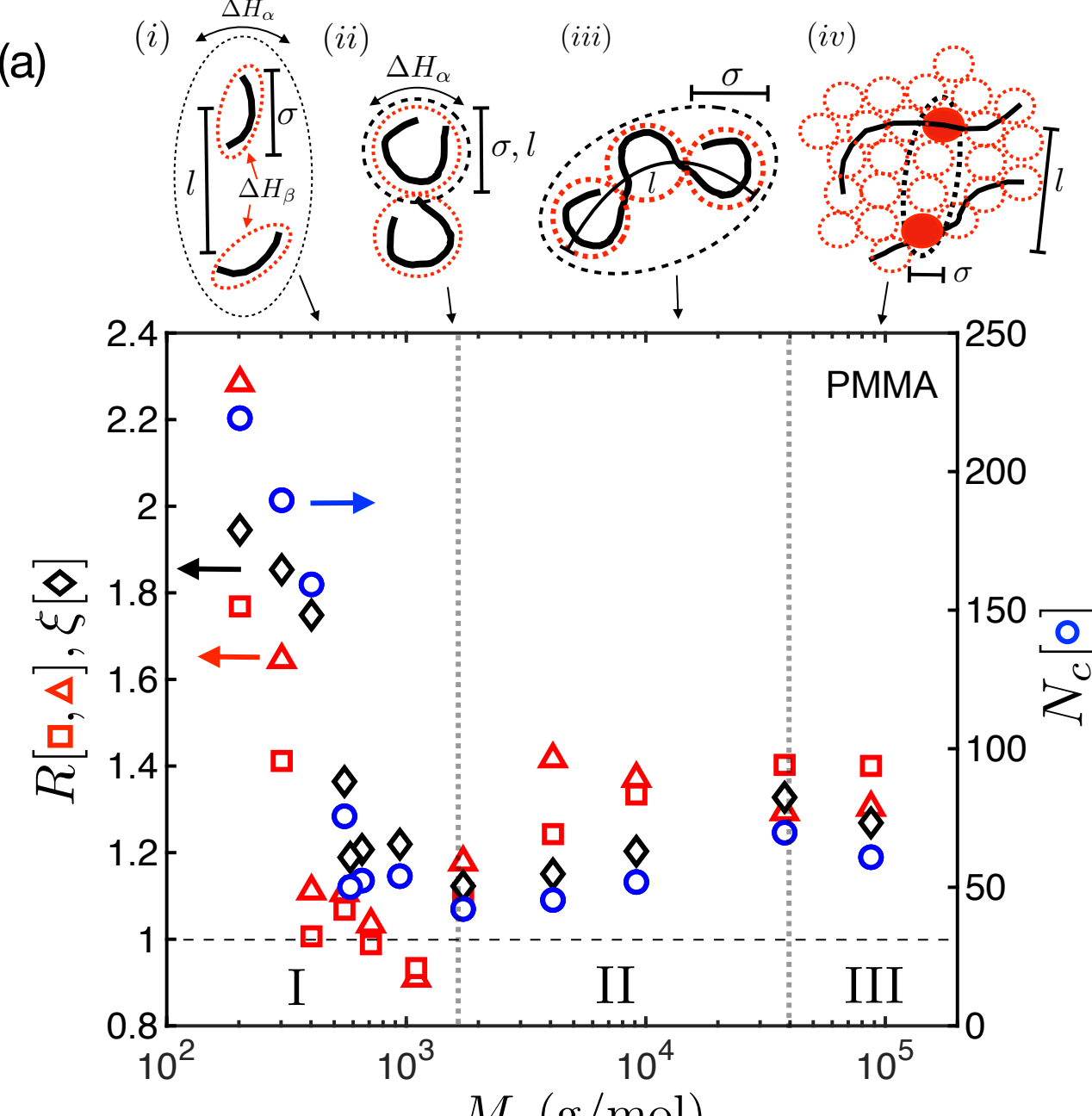

FIG. 3. Comparison between the molecular weight dependent ratio $\mathcal{R} = \Delta H_\alpha / \Delta H_\beta$ of $\alpha$ and $\beta$ activation enthalpies of PMMA (left axis, red squares and triangles), characteristic length scale $\xi_{\mathrm{DH}}$, (black diamonds) and the number of correlated monomers $N_c$ associated with the $\alpha$ relaxation (right axis, blue circles). Here, $\Delta H_\alpha(M) \equiv \Delta H_\alpha(M)|_{T=T_g}$ was estimated from $\tau_\alpha(T)$ using two different approaches (squares and triangles) following [29] and [43]. The dotted vertical lines mark the regime boundaries at $M^*$ and $M^{**}$. The cartoons (i-iv) illustrate the proposed relationship between the activation enthalpies for the $\alpha$ and $\beta$ relaxations within the different regimes and are described in the text.

Several recent computational studies have uncovered relations between DF and secondary relaxation contributions. Recent simulations by Guiselin *et al.* [58] and Scalliet *et al.* [59] used a swap Monte Carlo technique to access equilibrated temperatures near $T_g$, finding that slow regions relax by DF by spreading of mobility from more localised relaxations situated within an 'excess wing' on the high-frequency side of the structural $\alpha$ relaxation response (analogous to our picture of $\beta$ relaxations facilitating the $\alpha$ relaxation). Ortlieb *et al.* [22] combined simulations with experiments on a colloidal glass-former [22], finding that each 'particle' participating in a CRR takes part in many excitations (DF) during the life-time ($\sim \tau_\alpha$) of a CRR, leading the authors to speculate that CRRs form by accumulation of excitations [65]. Finally, Nishikawa and Berthier [60] demonstrated that, in a 3D lattice glass model near $T_g$, structural relaxation is driven by a small population of mobile particles (with low activation barriers) acting as emerging quasiparticles that drive DH.

*Conclusion*— We demonstrate how chain length and chain flexibility affect dynamic heterogeneities in linear polymers. Highly flexible polymers show simpler behavior with chain-length-independent dynamic heterogeneity, whereas less flexible polymers show a complex chain-length-dependence resulting from an interplay between intramolecular and intermolecular cooperativity. Our results provide a benchmark for new theories and models of glass-formation in polymers, and other glass-formers with internal degrees of freedom coupled to molecular motion.

*Acknowledgements*— We acknowledge the Engineering and Physical Sciences Research Council (EPSRC) for financial support (EP/M009521/1, EP/P505593/1, and EP/M506552/1). PDO thanks Georgetown University and the Ives Foundation for support. We thank Caroline Crauste for helpful discussions.

## END MATTER

We use two different methods for estimating the number of dynamically correlated units involved in the $\alpha$ relaxation. In the first approach due to Donth [11], the mean square temperature fluctuations $\delta T^2$ within a rearranging region are related to the breadth $\delta T$ of the calorimetric glass transition response in $c_p''$, and in turn to the volume $V_c$ of the correlated regions by $V_c = k_B T_\alpha^2 \Delta c_V^{-1}/(\rho\, \delta T^2)$ [3, 11].

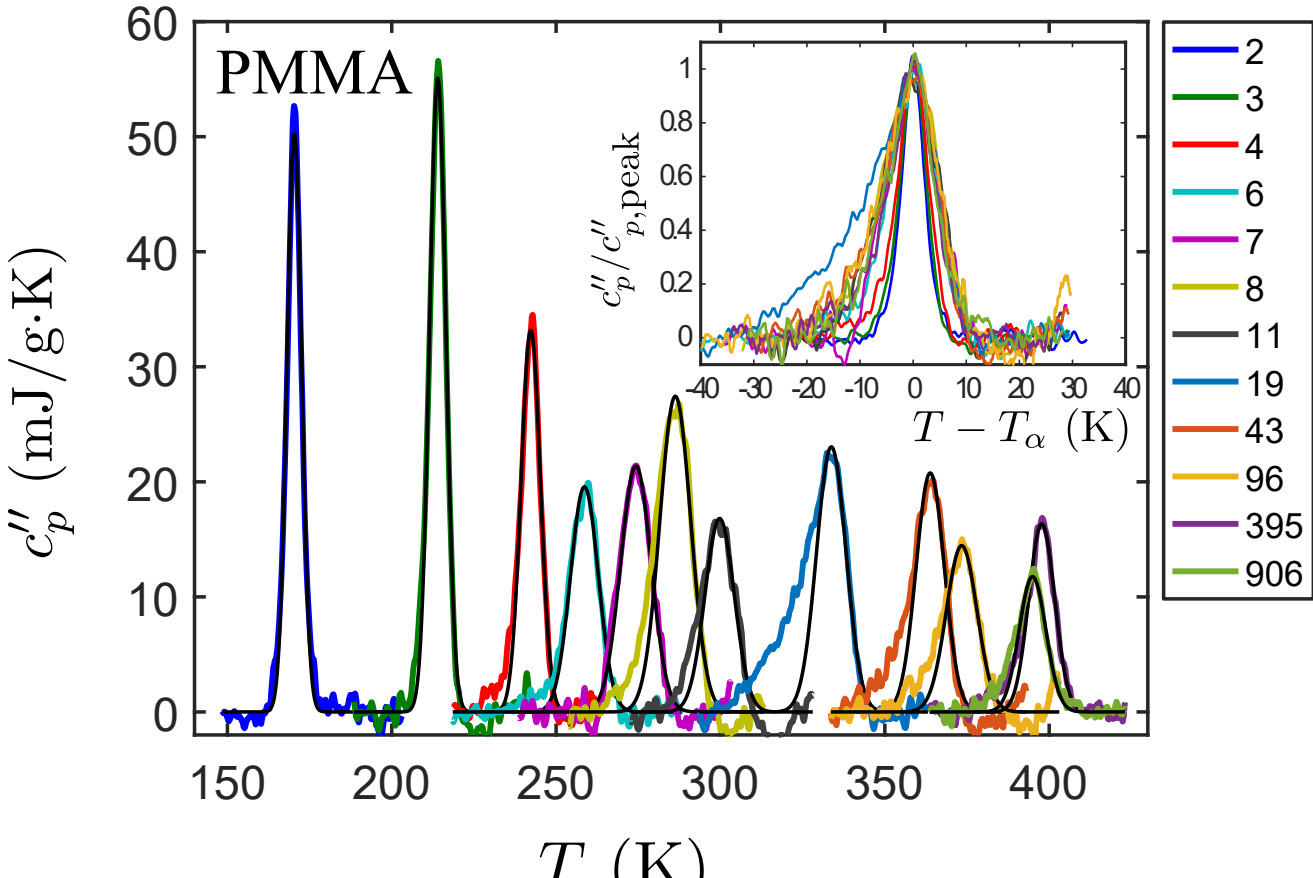

FIG. 4. The imaginary part of the $T$-dependent specific heat $c_p''(T)$ for PMMA. The legend shows the degree of polymerisation $n$. The peak in $c_p''(T)$ corresponds to the response due to the structural $\alpha$ relaxation, with Gaussian fits (black lines). The inset shows all $c_p''(T)$ data normalized and centered on the peak temperature $T_\alpha$.

The peak in $c_p''(T)$ occurs at $T_\alpha$, $\rho$ is the mass density, and $\Delta c_V^{-1} = c_{V,g}^{-1} - c_{V,l}^{-1}$ is the difference in reciprocal isochoric specific heat of the glassy and liquid states at $T_\alpha$. The number of monomers $N_c = \rho V_c N_\text{A}/M_0$ taking part in correlated motion can be estimated as:

$$N_c(M,P) = \frac{k_B N_\text{A} T_\alpha^2 \Delta c_V^{-1}}{M_0\, \delta T^2}, \quad (2)$$

where $N_\text{A}$ is Avogadro's number, $M_0$ is the monomer molar mass, and $P$ is the period of the TMDSC oscillation.

TMDSC measurements were performed on PMMA, PS, PPG-DME, and PDMS, as described in the Supplementary Material [43], yielding the complex specific heat capacity $c_p^*(T) = c_p'(T) - ic_p''(T)$. The real component $c_p'(T)$ has a step across the glass transition (see Fig. S1 of [43]), which yields $\Delta c_p^{-1}$. The imaginary component $c_p''(T)$ shows a peak at the transition temperature $T_\alpha$; $T_\alpha$ and $\delta T$ were calculated based on a Gaussian fit to $c_p''(T)$ as in [67]. We follow the literature [7, 9, 71] and approximate $c_V \approx c_p$, which slightly overestimates $N_c$ but does not affect our conclusions [72]

$c_p''(T;M)$ are shown for PMMA in Figure 4, measured using a modulation period $P = 60$s, corresponding to $\tau_\alpha \approx 10$s. A clear increase in the peak temperatures $T_\alpha$, and in the peak broadening $\delta T$ [73], are observed with increasing $M$. As shown in Fig. S5 (SI), the variation with molecular weight is smaller for the flexible polymers PPG-DME and PDMS, than for PMMA or PS, which is expected due the higher chain flexibility of PPG-DME and PDMS [29] that leads to a smaller variation in $T_g$.

The second method uses BDS to estimate the number of monomers within a dynamic correlation volume, using the fluctuation-dissipation-based approach of Berthier *et al.* [31, 32]. They showed that a "three-point" dynamic susceptibility is a lower bound to $\chi_4$, so that the number of dynamically correlated monomers $N_c^{(4)}(T)$ is

$$N_c^{(4)}(T) \approx \frac{k_B N_\text{A}}{m_0 \Delta c_P} T^2 \max_\omega \left\{ \left| \frac{\mathrm{d}\chi(\omega,T)}{\mathrm{d}T} \right| \right\}^2, \quad (3)$$

where $m_0 \Delta c_p$ is the difference in isobaric monomer molar heat capacity between the liquid and glass, and $\chi(\omega,T) = [\varepsilon'(\omega,T) - \varepsilon_\infty(T)]/[\varepsilon'(0,T) - \varepsilon_\infty(T)]$ is the normalized dynamic susceptibility [32]. Here, $\varepsilon'(\omega,T)$ is the real component of the complex permittivity and $\varepsilon_\infty(T)$ is its high-frequency limit. The structural relaxation times $\tau_\alpha(T)$ associated with the response $\chi(\omega,T)$ were determined using the fitting approach described in [43], which allowed conversion from $N_c^{(4)}(T)$ to $N_c^{(4)}(\tau_\alpha)$.

Fig. 5 shows $N_c^{(4)}(\tau_\alpha)$ for PMMA and PPG-DME from BDS data (open symbols), compared with $N_c$ calculated for $\tau_\alpha \approx 10$s from TMDSC data (solid symbols). In both cases, extrapolation of the BDS data shows that $N_c^{(4)}(10\text{s}) \simeq N_c$. For PMMA, $N_c^{(4)}(\tau_\alpha)$ was only calculated for oligomers ($n = 2$-7), since for longer chains the $\alpha$ relaxation response is increasingly obscured by a strong secondary $\beta$ relaxation [29], hindering an accurate determination of $\chi(\omega,T)$. The corresponding plots for PS and PDMS are shown in Fig. S7 of [43]. For PDMS we generally find good correspondence between $N_c^{(4)}(10\text{s})$ and $N_c$. For PS, a so-called "excess wing" [44, 70] on the high-frequency flank of the $\alpha$ relaxation obscures the $\alpha$ response [74], which increases the uncertainty in the absolute values of $N_c^{(4)}(\tau_\alpha)$. However, $N_c^{(4)}$ increases monotonically with $\tau_\alpha$, and both $N_c^{(4)}$ and $N_c$ show a similar variation with chain-length for $\tau_\alpha \sim 10\,\text{s}$.

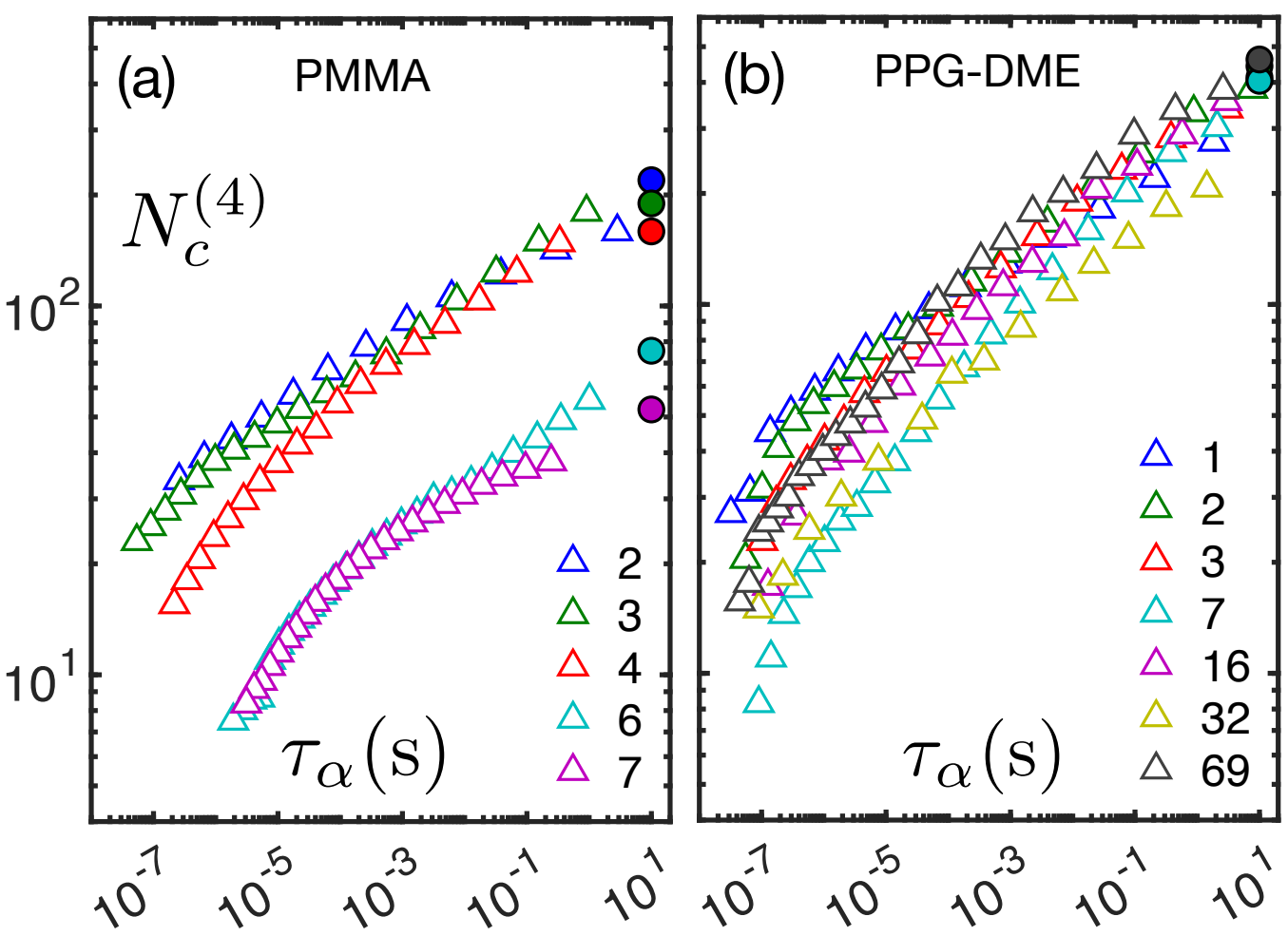

FIG. 5. The number of monomers $N_c^{(4)}$ ($\triangle$) undergoing correlated motion during the structural $\alpha$ relaxation for (a) PMMA and (b) PPG-DME, determined using BDS, as a function of the $\alpha$ relaxation time $\tau_\alpha$, for $n$ noted in the legends. The extrapolation of $N_c^{(4)}$ to $\tau_\alpha = 10\,\mathrm{s}$ agrees with the corresponding value $N_c(\tau_\alpha = 10\mathrm{s})$ from TMDSC ($\bullet$).

# Supplemental Information for:
# Chain-length-dependent correlated relaxation motion in polymers

Matthew Reynolds[1], Daniel L. Baker[1], Peter D. Olmsted[2], and Johan Mattsson[1]*
[1] *School of Physics and Astronomy, University of Leeds, Leeds LS2 9JT, United Kingdom and*
[2] *Department of Physics and Institute for Soft Matter Synthesis and Metrology, Georgetown University, Washington DC, 20057*

(Dated: November 16, 2025)

## A. TEMPERATURE-MODULATED DIFFERENTIAL SCANNING CALORIMETRY (TMDSC)

TMDSC measurements were performed using a TA Q2000 DSC with a liquid nitrogen cooling system. A sinusoidal heating/cooling profile with an amplitude of 1 K and a period of $P = 60$ s (corresponding to $\tau_\alpha = 60/2\pi \approx$ 10s) was superimposed onto an underlying cooling rate of $0.25 - 0.5$ K/min; an exception was made for PDMS for which the experiments were performed on heating following quenching to avoid sample crystallization. An additional necessary step for correcting the phase angle between heat flow and heating rate was carried out following the procedure outlined in Weyer *et al.* [1]. Using this methodology, the complex specific heat capacity, $c_p^* = c_p' - ic_p''$, was determined, where the real component $c_p'$ shows a step and the imaginary component $c_p''$ is manifested as a peak, as the structural $\alpha$ relaxation is probed.

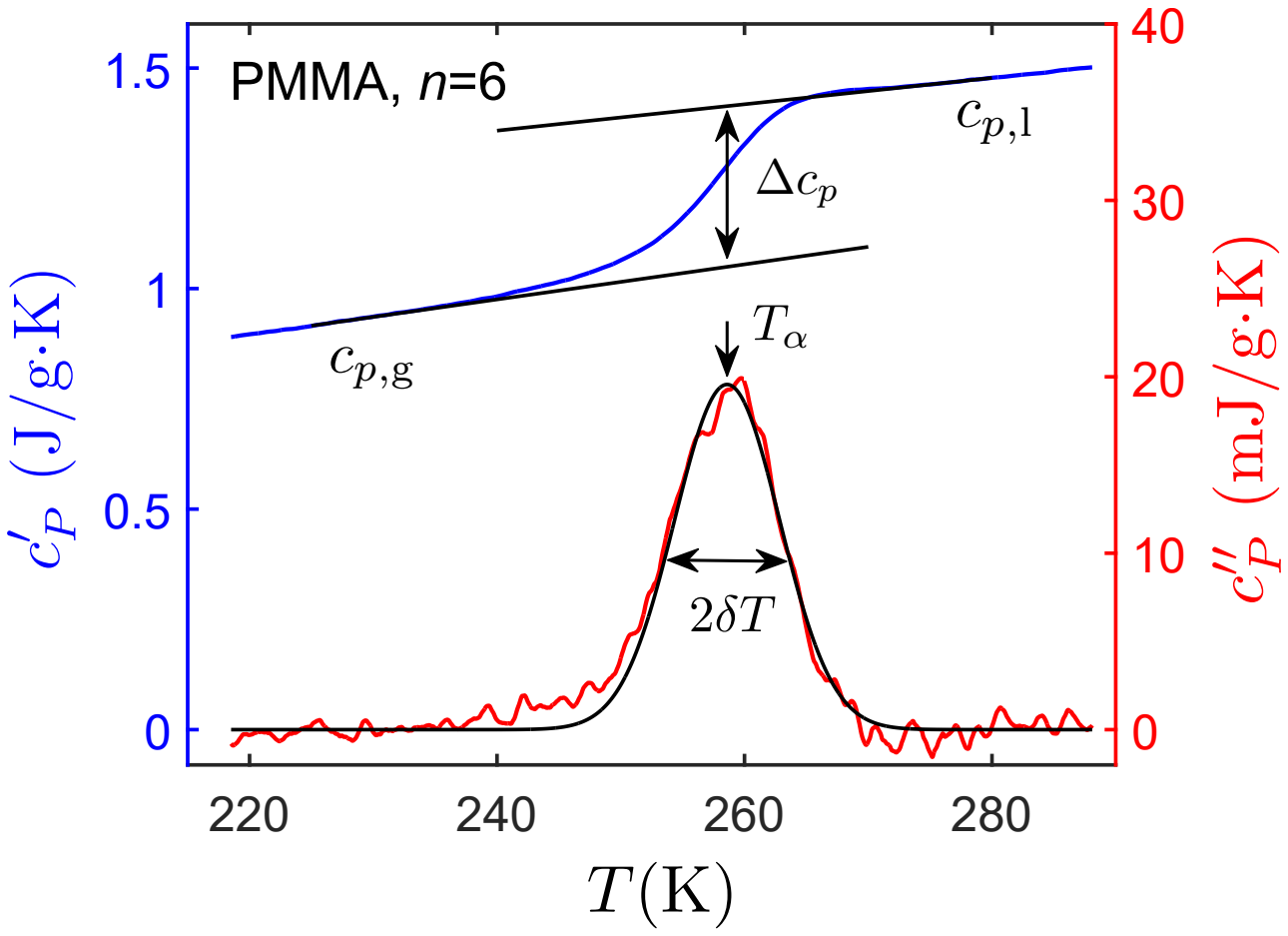

FIG. S1. TMDSC results for PMMA ($n = 6$) across the glass transition, using a modulation period of $P = 60$s and an amplitude of $A = \pm 1$K.

The results of a typical TMDSC measurement are shown in Fig. S1, where the real and imaginary components of $c_p^*$ are shown for PMMA with $n = 6$ ($M_W$=660 g/mol). $c_p'$ shows the typical step observed in a heat capacity measurement across the glass transition, where $\Delta c_p^{-1}$ can be determined from the change in the step in $c_p'(T)$, as the difference in reciprocal heat capacities of the glass and liquid states at $T_\alpha$ ($\Delta c_p^{-1} = c_{p,g}^{-1} - c_{p,l}^{-1}$)[2, 3]. $c_p''$ shows a peak, where $T_a$ and $\delta T$ can be determined from a fit to a Gaussian function:

$$c_p''(T) = \frac{A}{2\delta T\sqrt{\pi/2}} \exp\left[-\frac{1}{2}\left(\frac{T - T_\alpha}{\delta T}\right)^2\right]. \tag{S1}$$

The real component of the complex heat capacity is shown in Fig. S2 for (a) PMMA, (b) PS, (c) PPG-DME, and (d) PDMS, for different degrees of polymerisation, $n$. The step in $c_p'$ at $T_\alpha$ is akin to that observed in standard DSC, and the trend of increasing $T_\alpha$, and thus $T_g$, with $n$ is clearly observed.

The imaginary component of the specific heat $c_p''$ is shown in Fig. S3 for PMMA, PS, PPG-DME, and PDMS, respectively. The response of the structural ($\alpha$) relaxation manifests as a peak, where $T_\alpha$ corresponds to the peak temperature. The left-hand figures show the increase in $T_\alpha$ with increasing $n$, with Gaussian (equation S1) fits shown by black lines. The right-hand figure data are amplitude re-normalised and centered on the peak temperature $T_\alpha$. An increase in the breadth of the transition $\delta T$ for increasing $n$ is observed for all polymers. The observed asymmetry for longer chains is due to vitrification effects below $T_g$ [2], so the temperature range of the Gaussian fit was limited to $T \gtrsim T_g$. Fig. S4 shows the breadth $\delta T$ (left) and transition step height $\Delta c_P$ (right) as a function of molecular weight for the four polymer systems.

## B. BROADBAND DIELECTRIC SPECTROSCOPY (BDS)

BDS measurements were performed using a Novocontrol Alpha-A analyser with a Quatro Cryosystem, measuring the complex permittivity $\varepsilon^*(f)$ for frequencies $f$ within the range of $10^{-2}$ Hz $\leq \nu \leq 10^6$ Hz, with a temperature accuracy of $\pm$0.1K. To determine $N_c^{(4)}$ using Eq. 2 of the main paper, the determination of $T$-derivatives is required. To ensure that only contributions from the $\alpha$ relaxation are captured, thus minimising any contributions from additional secondary relaxations, the $\varepsilon^*(f)$

* k.j.l.mattsson@leeds.ac.uk

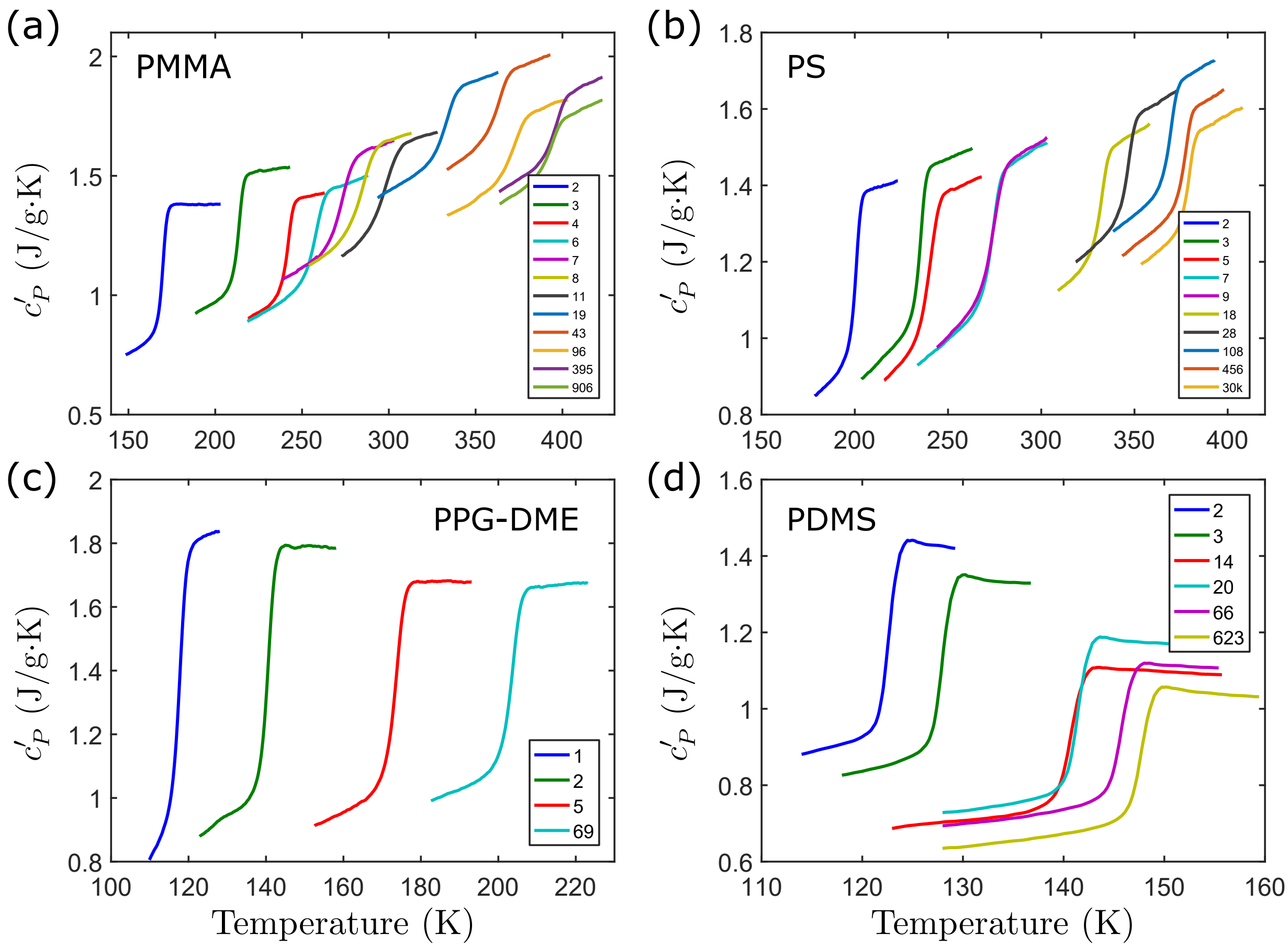

FIG. S2. The real part $c'_p$ of the specific heat response at $\omega = 2\pi/60\,\mathrm{s}^{-1}$ for the four polymers. The degrees of polymerization $n$ are shown in the legends.

spectra were fitted by a sum of relaxation contributions:

$$\varepsilon^*(\omega, T) = \frac{-i\sigma(T)}{\omega} + \varepsilon_\infty(T) + \sum_{j=1}^{N} \frac{\Delta\varepsilon_j(T)}{\left[1 + (i\omega\tau_j(T))^{m,j}\right]^{n,j}} \;, \quad \text{(S2)}$$

where $\omega = 2\pi f$, $\sigma$ is a fitted conductivity, and $\varepsilon_\infty$ is the high frequency permittivity. The relaxation contributions were described using Havriliak-Negami (HN) functions [4], where $N = 1, 2$ or 3, depending on how many relaxation modes $\alpha, \beta, \gamma$ are observed at a given temperature.

The renormalized susceptibility $\chi(\omega, T)$ is defined from the $\alpha$ relaxation contribution to $\varepsilon'(\omega, T)$ S2 by [5]:

$$\chi(\omega, T) = \frac{\varepsilon'(\omega, T) - \varepsilon_\infty(T)}{\varepsilon'(0, T) - \varepsilon_\infty(T)}. \quad \text{(S3)}$$

As an example, $\chi(\omega, T)$ for PDMS with $n = 2$, is shown in Fig. S5, where panel (a) shows the values of $\chi(\omega, T)$ as a function of angular frequency at different temperatures ranging from 121.5 to 128.5 K in steps of 0.5 K. From these data, the calculation of $T$-derivatives are necessary to calculate $N_c^{(4)}$, where the derivatives were determined using a finite difference approximation. The procedure used to determine $N_c^{(4)}$ follows that outlined by Dalle-Ferrier *et al.* [5]. $T|\mathrm{d}\chi(\omega, T)/\mathrm{d}T|$ (*i.e.* the $T$-derivative of $\chi(\omega, T)$ multiplied by $T$), is shown for PDMS ($n = 2$) in Fig. S5(b). An accurate determination of the peak maximum corresponding to each temperature is required to calculate $N_c^{(4)}(T)$ and Fig. S5(b) illustrates the effect of varying the $T$-differences used in the calculation of the corresponding derivatives, where each colour denotes a different $T$-step $\Delta T$ used in the calculation: $\Delta T$ =0.01 K (black); 1 K (red); 2 K (green); 4 K (blue). Clearly, small $T$-steps are required to determine accurate values of the peak maxima.

For our calculations we used $\Delta T = 0.01$K ($T \pm 0.005$K), corresponding to the black data in the figure. For each $T$, we know the $\alpha$ relaxation HN-parameters, corresponding to $\chi(\omega, T)$, from our fitting of the permittivity data. To determine the derivative corresponding to each $T$, we assume that the shape of the $\alpha$ relaxation ($\chi(\omega, T)$) does not change for the very small temperature differences used ($\Delta T = 0.01$K) and the derivative ($T|\mathrm{d}\chi(\omega, T)/\mathrm{d}T|$)

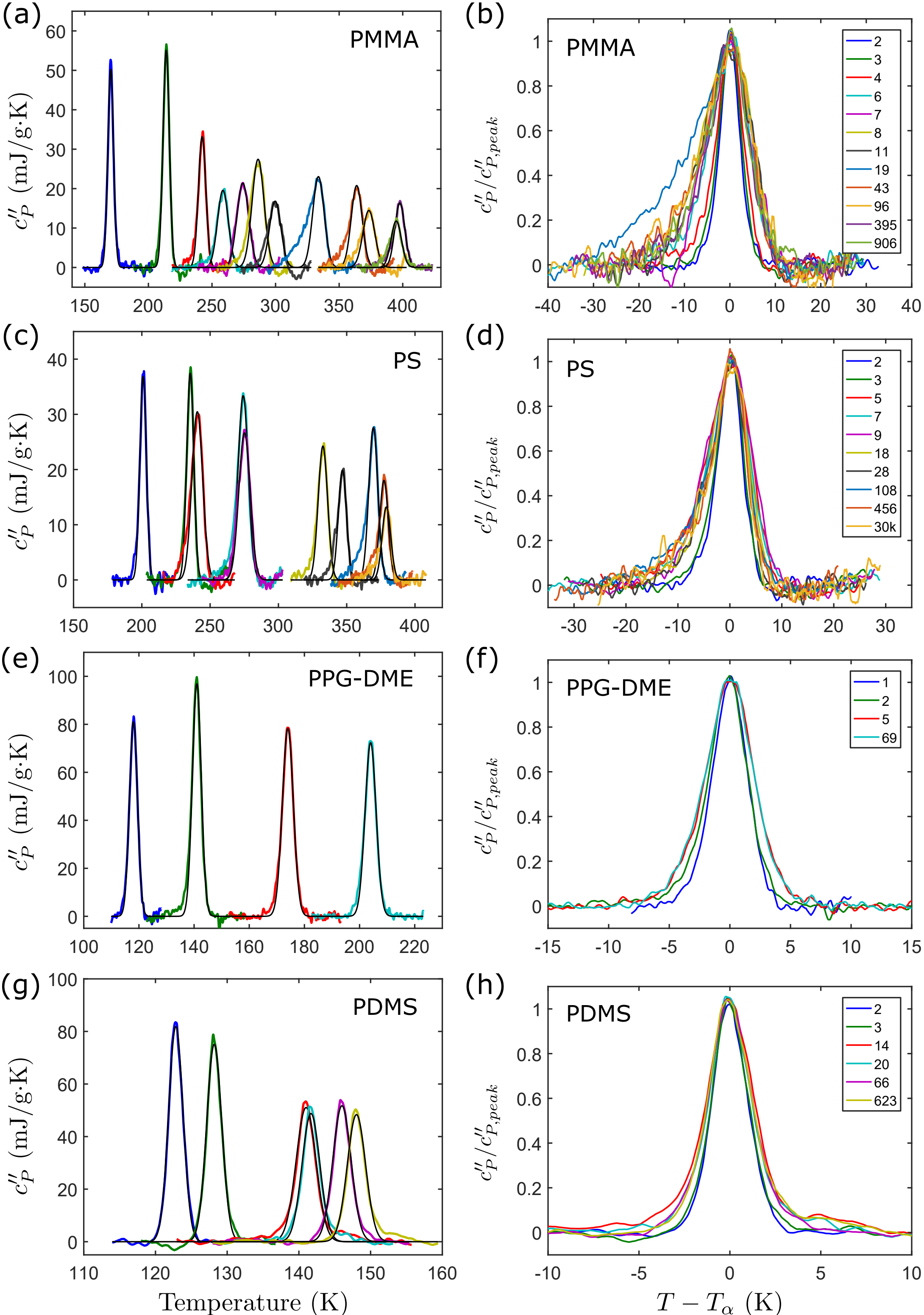

FIG. S3. (Left; a,c,e,g) $c_p''(T)$ ($P = 60$s) with Gaussian fits shown in black; (Right b,d,f,h) $c_p''(T)$ ($P = 60$s) show the data amplitude re-normalised and centered on the peak temperature $T_\alpha$, for PMMA, PS, PPG-DME, and PDMS, respectively. The legends show the degree of polymerisation $n$.

is thus set by the shape and size of the $\alpha$ relaxation response at each temperature $T$, and the difference in $\alpha$ relaxation time-scale ($\tau_\alpha(T)$) across $\Delta T$. For completeness, the $\alpha$ relaxation time-scale data $\tau_\alpha(T)$ are shown

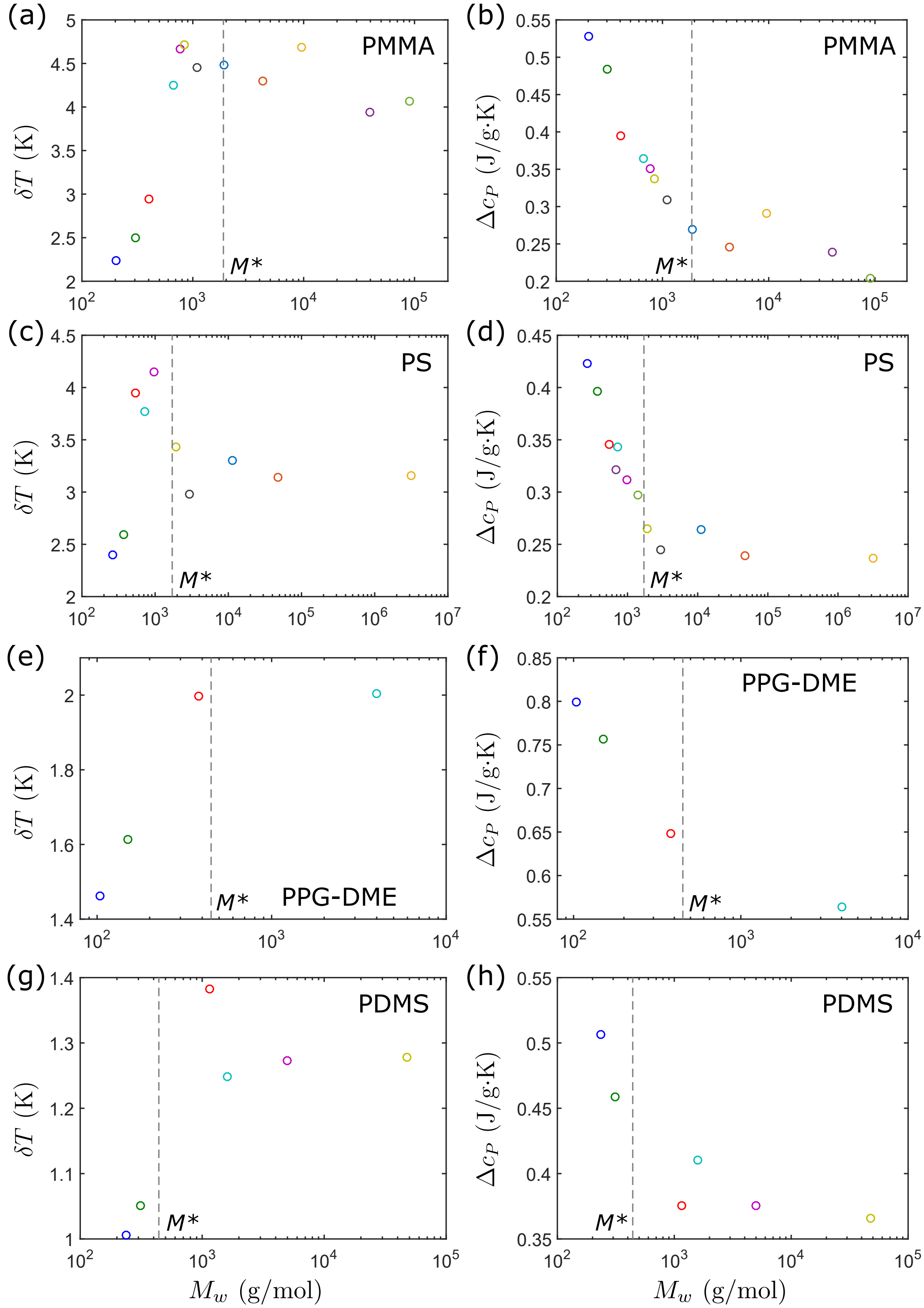

FIG. S4. (Left; `a,c,e,g`) Breadth of the peak in $c_p''$ as a function of molecular weight for PMMA, PS, PPG-DME, and PDMS. (Right, `b,d,f,h`) Step in the real part of the heat capacity $c_p'$, as shown in Fig. S1, for PMMA, PS, PPG-DME, and PDMS, as a function of molecular weight.

for all polymers and investigated $n$ and temperatures in an Arrhenius plot in Fig. S6.

## C. THE NUMBER OF DYNAMICALLY CORRELATED MONOMERS WITHIN THE $\alpha$ RELAXATION

Fig. S7 shows the number of dynamically correlated monomers $N_c^{(4)}(\tau_\alpha)$ (open symbols) for PMMA, PS,

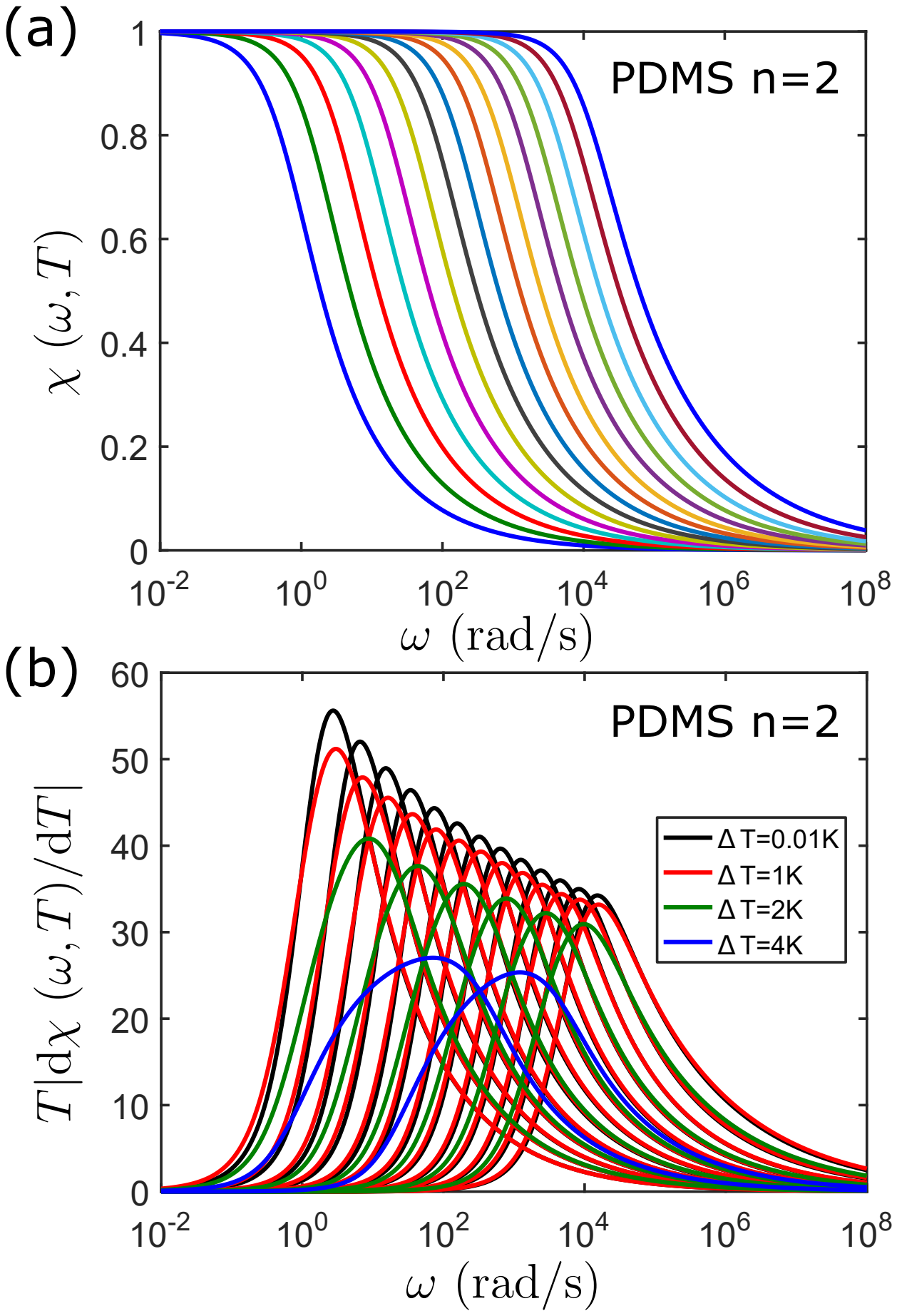

FIG. S5. (a) Normalised susceptibility (calculated using Eq. S2 from Havriliak-Negami $\alpha$ relaxation parameters) for PDMS (polymerization $n = 2$) as a function of angular frequency $\omega$. The different colors correspond to different temperatures, ranging from 121.5 to 128.5 K in steps of 0.5 K. (b) $T|\mathrm{d}\chi(\omega, T)/\mathrm{d}T|$ versus $\omega$ for PDMS ($n = 2$). The different colors represent different temperature differences used to compute the derivatives, as described in detail in the text.

PPG-DME, and PDMS, respectively. $N_c^{(4)}(\tau_\alpha)$ are calculated from BDS data, as described in the main paper. Also, for each polymer and chain-length, the solid symbol shows the number of dynamically correlated monomers $N_c$ as determined from TMDSC measurements at a period of 60 s, corresponding to a relaxation time $\tau_\alpha = 10\,\mathrm{s}$. The extrapolation of $N_c^{(4)}(\tau_\alpha)$ to $\tau_\alpha = 10\,\mathrm{s}$ generally agrees well with $N_c$, even though for PS the existence of a so-called "excess wing" [6, 7] on the high-frequency flank of the $\alpha$ relaxation, measured using BDS, makes the comparisons between the two approaches more difficult. For all four polymer systems, $N_c^{(4)}$ increases monotonically with $\tau_\alpha$, and both $N_c^{(4)}$ and $N_c$ show a similar variation with chain-length for $\tau_\alpha \sim 10\,\mathrm{s}$.

## D. COMPARISON BETWEEN THE ACTIVATION BARRIER RATIO AND THE LENGTH-SCALE OF CORRELATED MOTIONS

If one naively assumes that the $N_c$ dynamically correlated monomers responsible for the $\alpha$-relaxation are distributed homogeneously within a spherical domain of size $\xi$, with some volume fraction $\phi$ (generally $\phi < 1$), then we expect a scaling

$$\xi = \left(\frac{3N_c V_0}{4\pi\phi}\right)^{1/3}, \tag{S4}$$

where $V_0$ is the monomer volume. This simple argument does not consider a more complex internal structure of the correlation volume, such as a fractal interior. As discussed in the main paper, both computer simulations and experiments [8–11] have studied the $T$-dependent geometry of regimes of correlated motions, often observing that dynamic clusters become increasingly compact near $T_g$ so that the characteristic length-scale $\xi \propto N_c^\gamma$ with $\gamma \approx 1/3$. Thus, in Fig. S8 we compare the length-scale $\xi = aN_c^{1/3}$ with the enthalpy barrier ratio $\mathcal{R} = \Delta H_\alpha/\Delta H_\beta$ for PMMA, and show that these two quantities scale similarly as a function of molecular weight. Note that $\Delta H_\alpha(M) \equiv \Delta H_\alpha(M)|_{T=T_g}$ was estimated from $\tau_\alpha(T)$ using two different approaches (red squares and triangles) following [4]. The data shown in red squares were determined from $\tau_\alpha(T_g) = \tau_0 \exp\left[\Delta H_\alpha(T_g)/RT_g\right]$ by setting $\tau_0 = \tau_0^{\mathrm{micr}} = 10^{-12}\mathrm{s}$, where $\tau_0^{\mathrm{micr}}$ is a microscopic time-scale. The data shown in red triangles were determined by equating $\tau_\alpha(T_g)$ from the Arrhenius relation to $\tau_\alpha(T_g) = \tau_0 \exp\left[(DT_0/(T_g - T_0)\right]$, where $\tau_0$, $D$, and $T_0$ are VFT fitting parameters, which results in $\Delta H_\alpha \equiv DT_0 RT_g/(T_g - T_0))$ [4].

## E. THE BARRIER BEHAVIOUR IN POLYMERS

**PS and PAMS compared with PMMA**— Very little data exist in the literature for the $M$-dependence of the secondary $\beta$ relaxation barriers in polymers. In Figure 5 of Ref. [4], a combination of our data and literature data for $\Delta H_\beta(M)$ were presented for PMMA and polybutadiene (PB); however, for PB the existing data are limited to $M \gtrsim M^*$ which prevents a comparison of the $M$-dependence of the barriers across the three regimes. Bershtein *et al.* [12, 13] have reported $\Delta H_\beta(M)$ data for polystyrene (PS) and poly($\alpha$-methyl styrene) (PAMS), as shown in Fig. S9. They use a somewhat unconventional experimental approach to detect the $\beta$ relaxation using DSC, whereby the samples were quenched from the melt into the glass, annealed at a temperature within the glass, followed by heating runs at varying rates. Following this procedure, an enthalpy relaxation peak was observed that, based on its temperature location, was assigned as the $\beta$ relaxation, so that the activation barrier $\Delta H_\beta(M)$ was determined from its rate dependence. For PS and PAMS, the results are qualitatively similar to those observed for PMMA, i.e. a strong increase in $\Delta H_\beta(M)$ for short $M$, and $M$-independent behaviour for larger $M > M^*$; see Fig. S9 for a direct comparison between the three polymers.

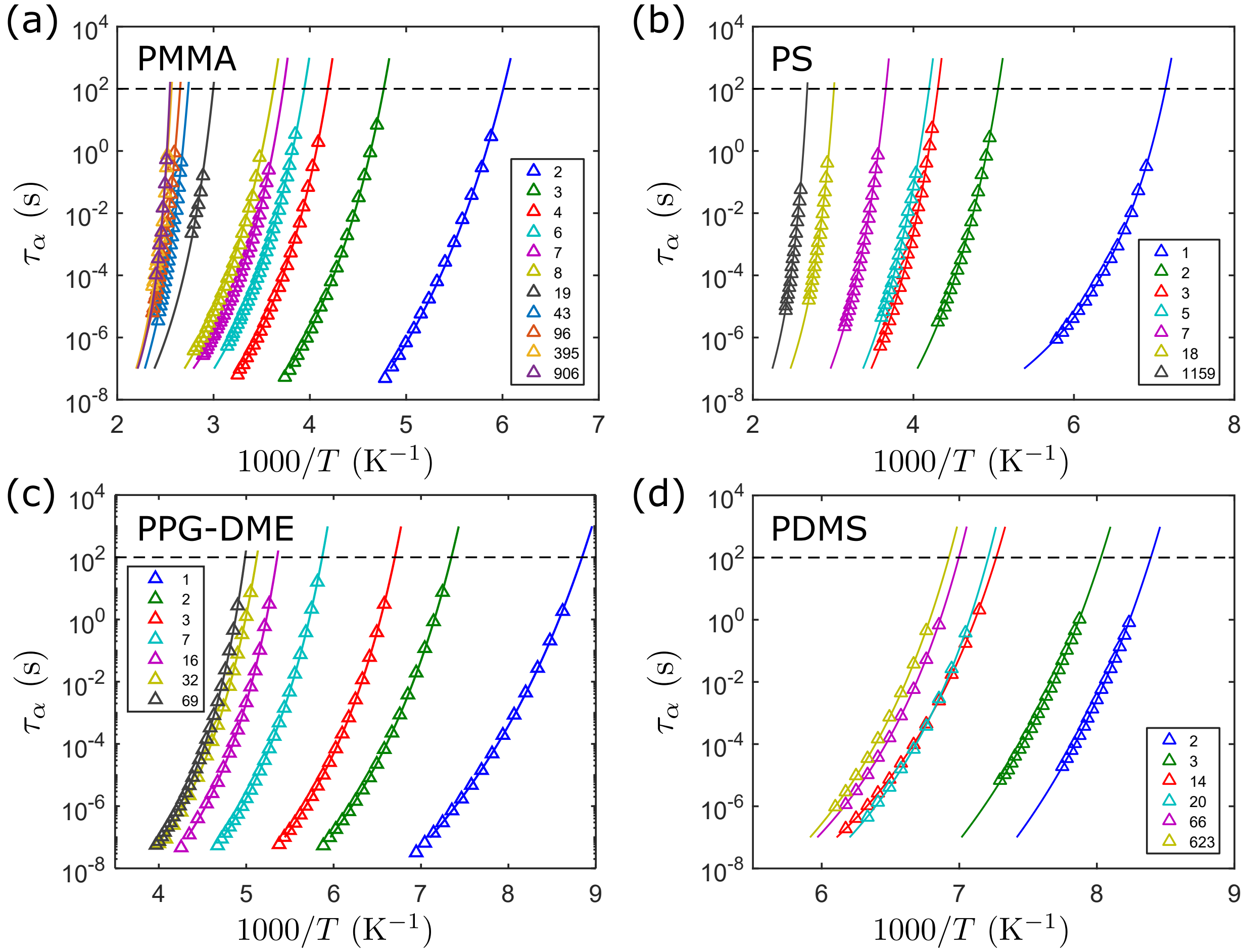

FIG. S6. Arrhenius plots for (a) PMMA, (b) PS, (c) PPG-DME, and (d) PDMS. The solid lines are fits to Vogel Fulcher Tammann (VFT) expressions. The dashed lines mark $\tau_\alpha$=100 s, since $T_g \equiv T(\tau_\alpha = 100\,\text{s})$. The legend shows the degrees of polymerisation $n$.

**PC**—Bershtein et al. [12, 13] also reported data for polycarbonate (PC), but these are more difficult to interpret since two separate relaxations were detected and only a few data points were measured. Despite the scant data, the general trend of each of these components is similar to that observed for PMMA, PS, and PAMS [4], in that an increase in activation is observed for increasing $M$, followed by a flattening out towards an $M$-independent behaviour, where the transition takes place for $M \sim M^*$. For PC, the $\beta$ relaxation is generally very weak and often not resolvable [14–16], and the more easily measurable faster $\gamma$ relaxation for PC includes several relaxation contributions [14, 15]. Due to these uncertainties we are not confident in using the Bershtein data for PC.

**PDMS**—For poly(dimethyl siloxane) (PDMS), no distinct $\beta$ relaxation has, to the best of our knowledge, been reported in the literature. However, instead of a distinct $\beta$ relaxation, PDMS shows a so-called excess wing, a high-frequency flank on the structural $\alpha$ relaxation peak (discussed in more detail below); the excess wing for PDMS is clearly observed in dielectric spectroscopy data across the full molecular weight range [7]. Bershtein *et al.* [12, 13] reported $\Delta H(M)$ data for PDMS that they interpreted as arising from a $\beta$ relaxation. However, the barrier data show a smooth $M$ variation strongly resembling that of $\Delta H_\alpha(M)$ (including its associated excess wing) [4], and it is thus not clear what relaxation component the Bershtein data for PDMS actually probe. In summary, to the best of our knowledge, the only literature $\Delta H_\beta(M)$ data to directly compare with our PMMA data are those for PS and PAMS [12, 13].

**The barrier ratio $\mathcal{R}$ for PS and PAMS compared with PMMA**— To estimate the barrier ratio $\mathcal{R}$ for PS and PAMS, $\Delta H_\alpha(M)$ was first determined directly from the $T_g(M)$ literature data [4]. Since $\tau_\alpha(T_g) \equiv$ $100\ \text{s} = \tau_0 \exp\left[\Delta H_\alpha(T_g)/RT_g\right]$, we can directly estimate $\Delta H_\alpha(T_g)$ by setting $\tau_0 = \tau_0^{\text{micr}} = 10^{-12}$ s [4], where $\tau_0^{\text{micr}}$ is a microscopic timescale, so that $\Delta H_\alpha(T_g) = 14RT_g \ln 10 \simeq 32.2RT_g$. In addition, we use the literature $\Delta H_\beta(M)$ data from Bershtein *et al.* for PS and PAMS [12, 13], as shown in Fig. S9, and fit the data to the form $\Delta H_\beta \sim \ln M$ for $M < M^*$, followed by a plateau for $M > M^*$. To estimate the barrier ratio $\mathcal{R}$, we divide $\Delta H_\alpha(M)$ for PS and PAMS, respectively, by the fit to the $\Delta H_\beta(M)$ data for the combined PS and PAMS data. The resulting $\mathcal{R}(M)$ are shown in Fig. S10) where the abscissa has been normalised by $M^*$ [4] and the ordinate is normalised by its high-$M$ value to facilitate a direct comparison of the $M$-dependence between the two polymer systems. We find that both PS and PAMS demonstrate regime behaviour similar to that observed for PMMA: for short $M \leq M^*$ the ratio $\mathcal{R}$ falls steeply with increasing $M$, whereas for $M \gg M^*$ the ratio $\mathcal{R}$ increases weakly towards a plateau.

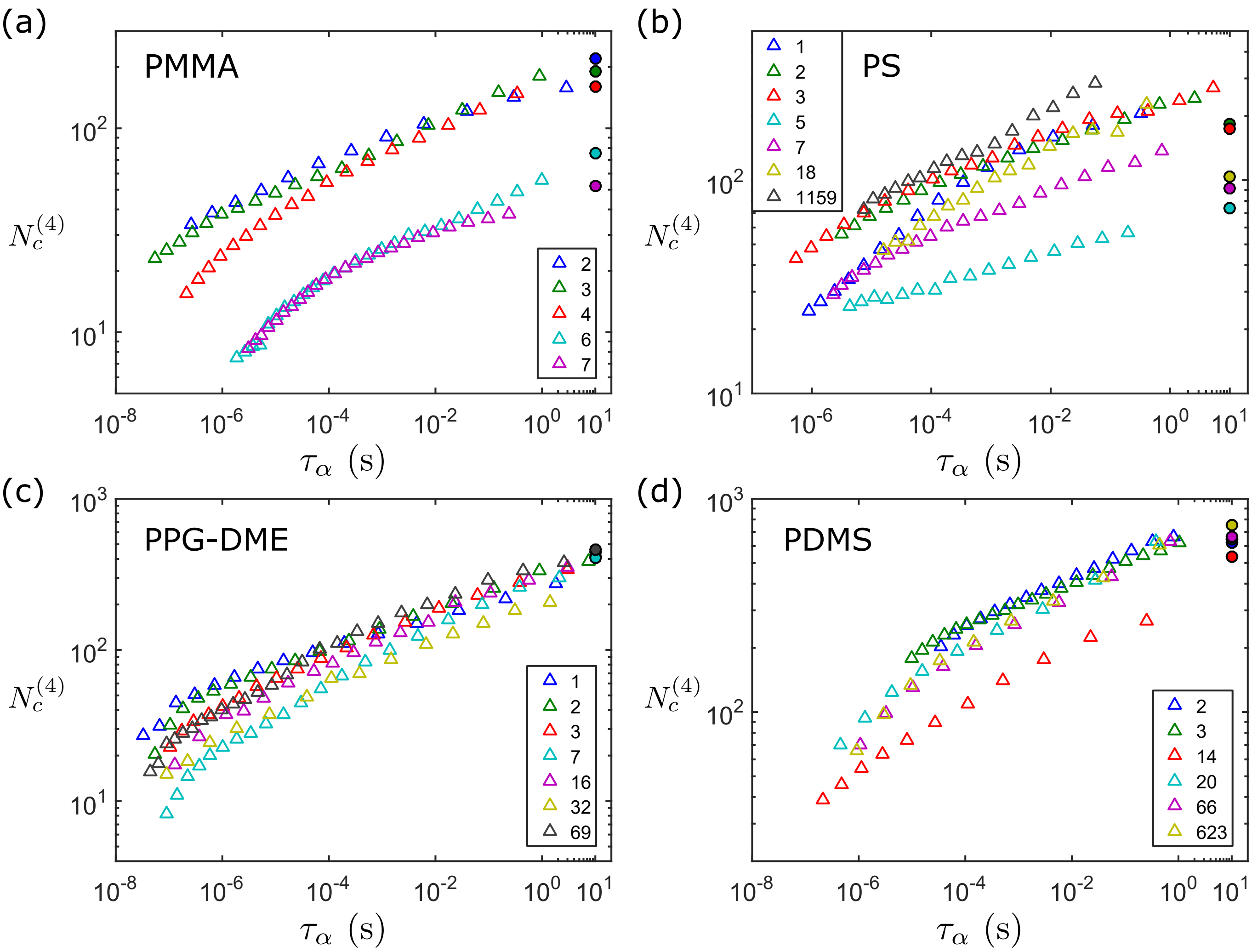

FIG. S7. The number of monomers $N_c^{(4)}(\triangle)$ undergoing correlated motion during the structural $\alpha$ relaxation for (a) PMMA and (b) PS, (c) PPG-DME, and (d) PDMS, determined using BDS, as a function of the $\alpha$ relaxation time $\tau_\alpha$. The sample degrees of polymerisation $n$ are listed in the legends. Generally, the extrapolation of $N_c^{(4)}$ to $\tau_\alpha = 10\,\text{s}$ agree with the corresponding value $N_c(\tau_\alpha = 10\text{s})$ from TMDSC ($\bullet$), as described in detail in the text.

**'Excess wing' in PDMS and PPG-DME**— As discussed above, to the best of our knowledge no $\beta$ relaxation has been directly reported in the literature for PDMS. However PDMS demonstrates a clear 'excess wing' for $M$ ranging from short oligomers to long entangled polymers [7]. Two main interpretations of excess wings exist in the literature: (i) it is an intrinsic feature of all $\alpha$ relaxations but is difficult to observe for many glass-formers [17, 18]; or (more commonly) (ii) it reflects a secondary (Johari-Goldstein) $\beta$ relaxation that is merged with the $\alpha$ relaxation [17]. The former interpretation has mainly been argued based on scaling arguments for the response function [6, 19], but many of these arguments were later questioned [20–22]. The latter interpretation, on the other hand, is supported by the fact that in many systems the separation between the $\alpha$ and $\beta$ relaxations can be increased (*e.g.* by aging experiments [23]; application of pressure [17]; 'confinement' within a matrix with more restricted dynamics, such as a clay [24] or a polymer matrix [25]; or studies of binary blends [26]). Under these conditions the excess wing becomes a separate relaxation contribution (a peak in the loss response). Moreover, recent computer simulations on binary spheres with Lennard-Jones interactions probed near $T_g$ using swap Monte-Carlo simulations [18, 27] demonstrated an excess wing relaxation contribution. It was found that this excess-wing contribution acted as a pre-cursor to the structural $\alpha$ relaxation, as mediated by the mechanism of dynamic facilitation. Thus, there is a strong case supporting that the excess wing reflects the presence of a secondary relaxation contribution that has effectively merged with the $\alpha$ relaxation [17]. We note that highly flexible polymers, such as PPG-DME and PDMS, have relatively low $T_g$ values, resulting in the $\alpha$ and (typically) more 'local' $\beta$ relaxation to be characterised by relatively similar time-scales near $T_g$; this could result in 'merging' of the two relaxation contributions.

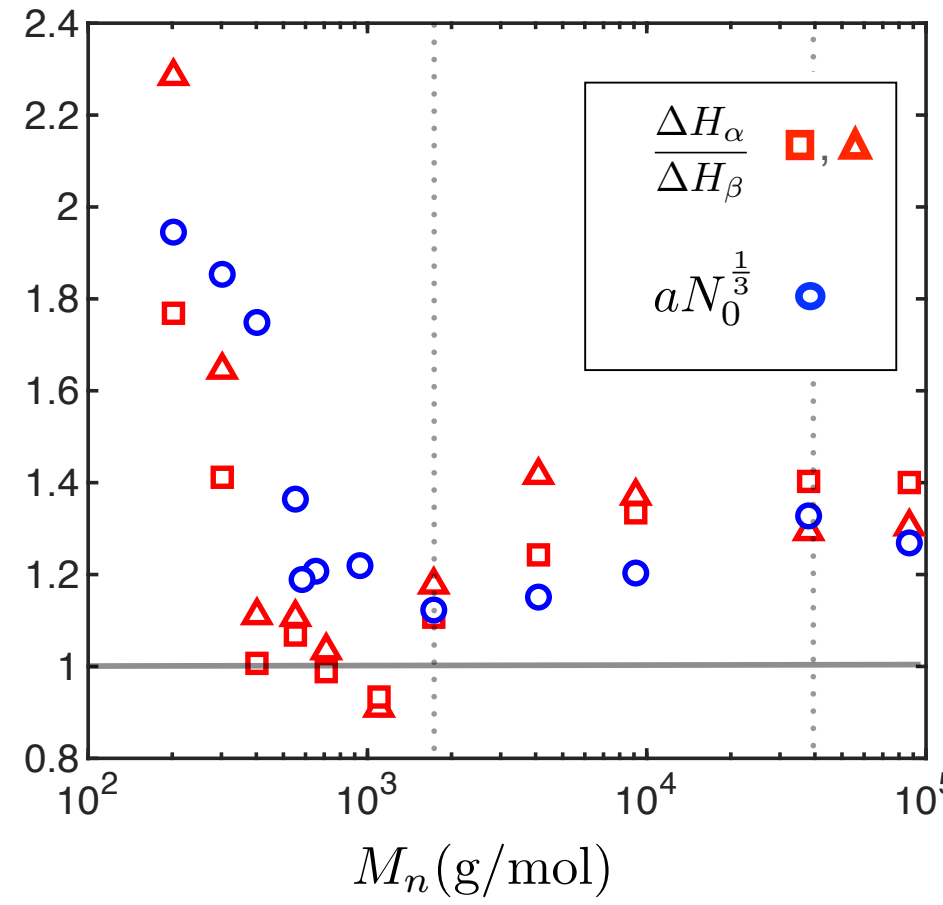

FIG. S8. A demonstration, for PMMA, that the ratio of the activation barriers $\mathcal{R} = \Delta H_\alpha / \Delta H_\beta$ for the $\alpha$ and $\beta$ relaxation, respectively, and the lengthscale characterising correlated dynamics $\xi$ scale similarly with molecular weight, given the assumption $\xi \sim N_c^{1/3}$, as discussed in the text.

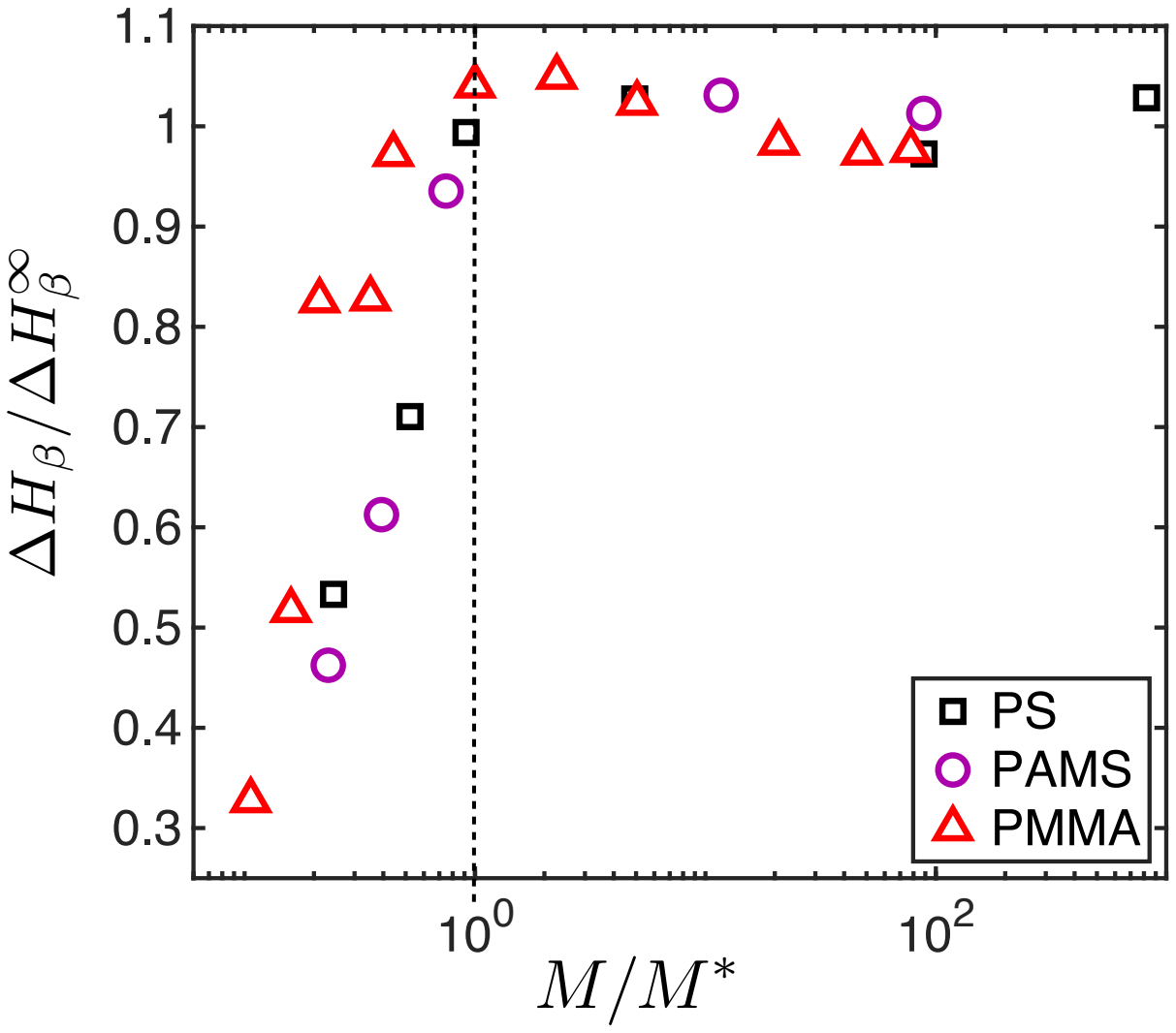

FIG. S9. The activation enthalpy of the $\beta$ relaxation ($\Delta H_\beta$) renormalised by its high-$M$ value ($\Delta H_\beta^\infty$), as a function of molecular weight $M$, renormalised by the molecular weight $M^\star$ at which the polymer dynamics and structure change from regime I to regime II (see discussion in the main paper).

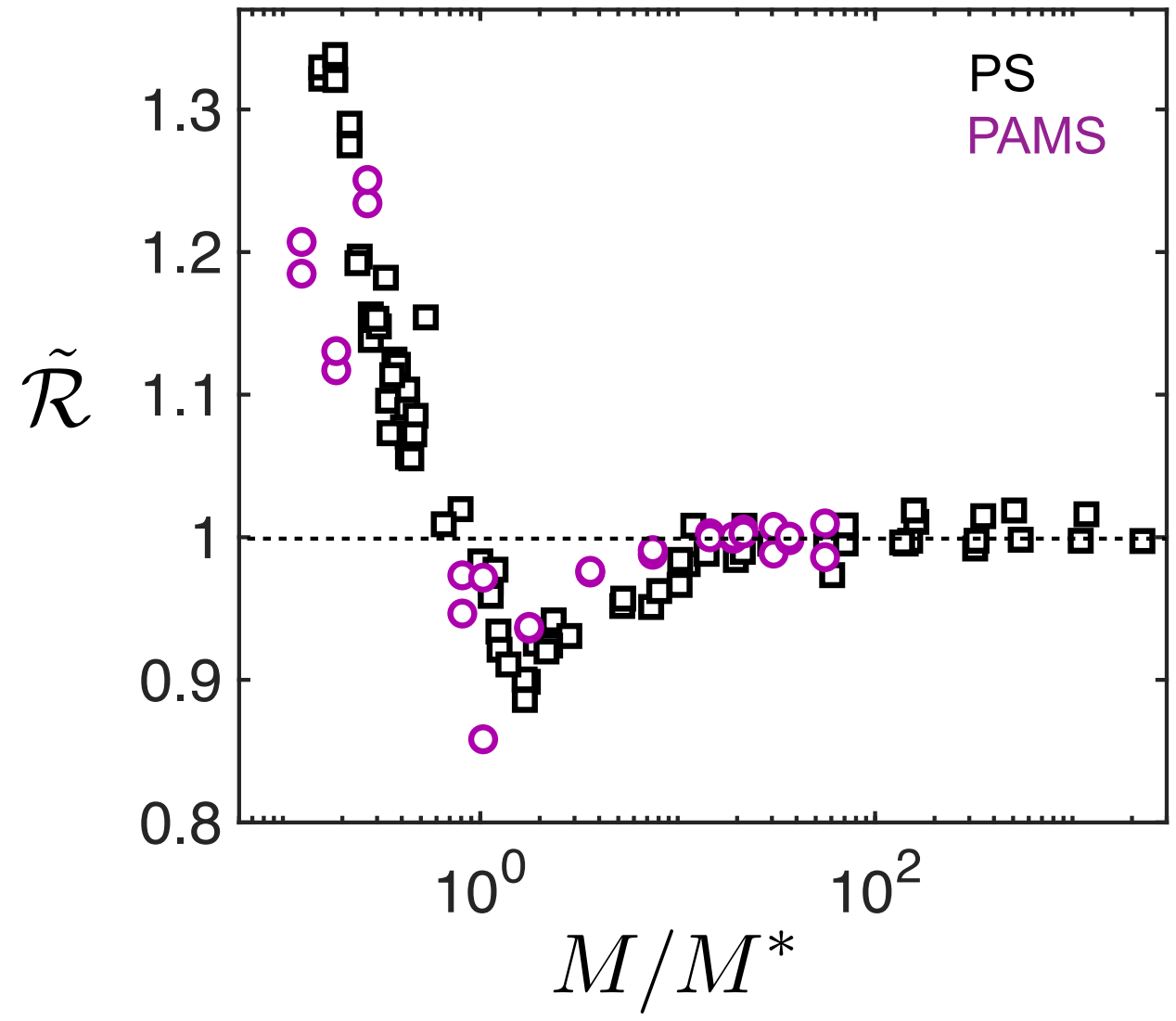

FIG. S10. The molecular weight-dependent barrier ratio $\mathcal{R}(M) = \Delta H_\alpha(M)/\Delta H_\beta(M)$ for polystyrene (PS; black squares) and poly(alpha methyl styrene) (PAMS; magenta circles). For direct comparisons, the barrier ratio is shown as $\tilde{\mathcal{R}} \equiv \mathcal{R}(M)/\mathcal{R}(\infty)$, normalised by its long-chain value, as function of the molecular weight $M$ normalised by $M^*$, characterising the transtion from regime I to regime II.

For PPG-DME (and PPG), the commonly observed secondary relaxation has a relatively low activation enthalpy of $\sim$ 32 kJ/mol [28, 29] and importantly does not show the pressure-dependence that is typical of (Johari-Goldstein) $\beta$ relaxations [17, 30]. Thus, the secondary relaxation in these systems has sometimes been referred to as a $\gamma$ relaxation [31, 32] and been assigned to a more 'local' nature. This assignment suggests, however, that an additional secondary $\beta$ relaxation might exist between the $\alpha$ and $\gamma$ relaxations. Broadband dielectric spectroscopy (BDS) investigations have indeed identified an additional secondary relaxation [29] in PPG, even though it has also been demonstrated that this relaxation is related to the common presence of a small amount of water since PPG and PPG-DME are hygroscopic [29, 31–33]. Consequently, the strength of this contribution decreases dramatically upon drying, and it was demonstrated that the contribution at atmospheric pressure can be removed by careful drying (*e. g.* using freeze-thaw cycling and/or heating under vacuum) [28, 31, 33–35]. Thus, from ex-periments performed at atmospheric pressure it was unclear whether dry PPG and PPG-DME have an additional secondary relaxation that is dielectrically just too weak to detect in the dry samples. However, in high-pressure studies of dry long-chain PPG (long enough for hydrogen-bond mediating chain-ends to have minimal influence) [31], the $\alpha$ relaxation is shifted to longer times, thus widening the dynamic window between the $\alpha$ and the readily observed pressure insensitive secondary relaxation (we refer to this secondary relaxation as the $\gamma$ relaxations in the following), making it easier to search for the presence of additional relaxation contributions. Indeed, such studies of PPG at high-pressures [31] found evidence for the presence of an excess wing, suggesting that PPG-DME also has an excess wing, and thus a merged $\beta$ relaxation, just like PDMS (within the interpretation that the excess wing is caused by an underlying 'merged' $\beta$ relaxation).

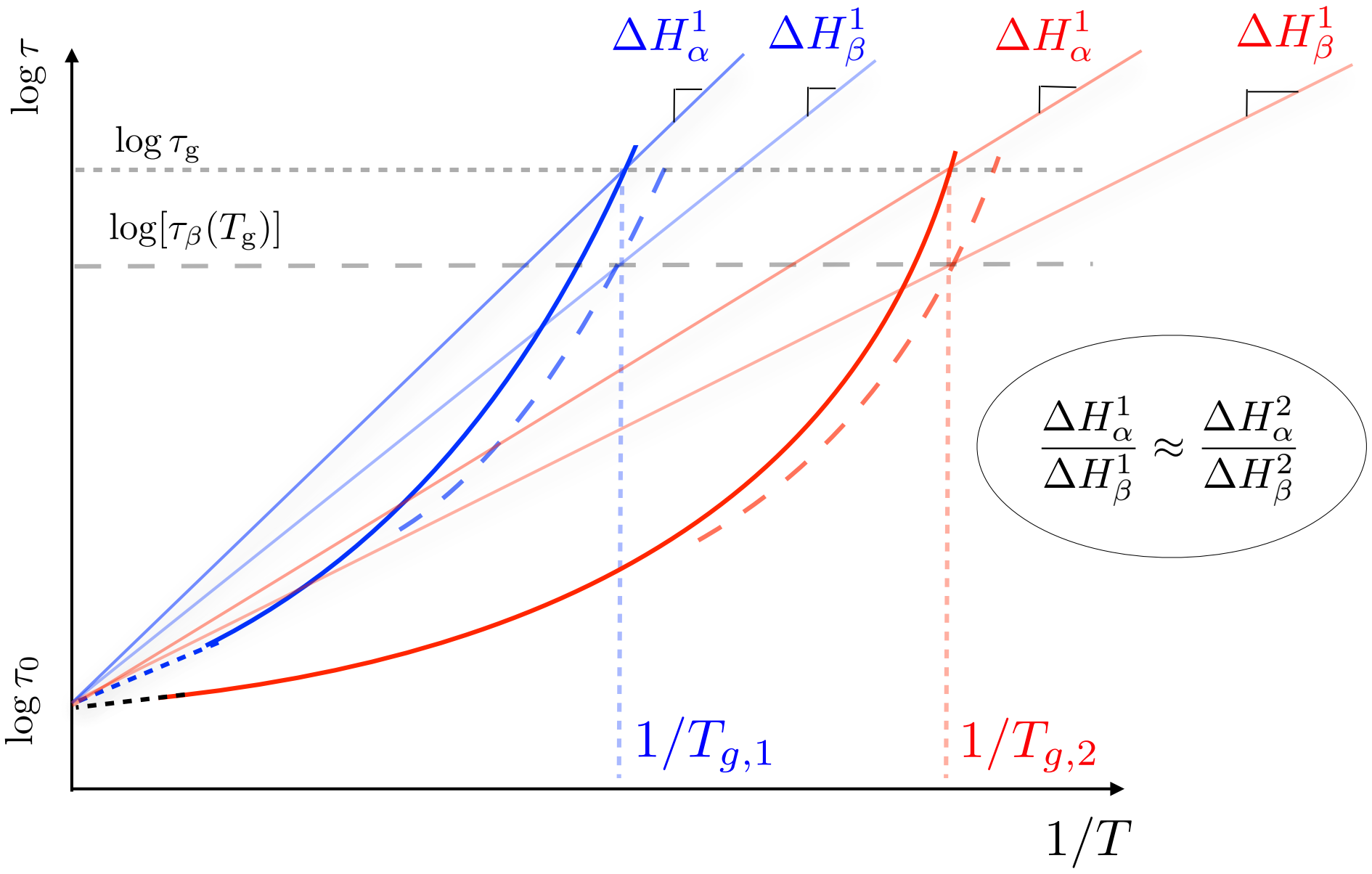

FIG. S11. Schematic illustrating the determination of the barrier ratio $\mathcal{R} = \Delta H_\alpha/\Delta H_\beta$ for a polymer with relatively high molecular weight $M$, and thus a high $T_g$ (blue), and the same polymer with lower $M$, and thus lower $T_g$ (red). The thick solid lines represent the $T$-dependent non-Arrhenius $\alpha$ relaxation time-scale ($\tau_\alpha(T)$) and the thick dashed lines represent a temperature-dependent characteristic time-scale for the $\beta$ relaxation ($\tau_\beta(T)$) that yields the 'excess wing' on the high-frequency side of the $\alpha$ relaxation (as observed for PDMS and PPG-DME).

**Barrier Ratio $\mathcal{R}$ for PDMS and PPG-DME**— If both PPG-DME and PDMS are characterised by merged $\alpha$ and $\beta$ relaxations, the key question is what this means for the corresponding barrier ratio? To better understand this, Fig. S11 schematically illustrates the determination of activation barriers for relaxation contributions in a glass-forming system. The thick blue solid line illustrates the $\alpha$ relaxation for a glass-former with a relatively high $M$, and thus a high $T_g$, while the thick red solid line illustrates the $\alpha$ relaxation for the same polymer, but with a lower $M$ and thus a lower $T_g$. The corresponding thick dashed lines indicate the approximate position of the 'merged' secondary $\beta$ relaxation contribution corresponding to the excess wing, situated on the high-frequency side of the $\alpha$ relaxation distribution, and thus on the low-$T$ side, for a fixed $\tau$ value.

Without a detailed model it is difficult to identify the secondary relaxation distribution from experimental relaxation data. However, the contribution of the merged relaxation contribution should be situated near the $\alpha$ relaxation, but be characterised by shorter time-scales, as marked by the dashed coloured lines indicating their position. The grey horizontal dashed lines mark the time-scale $\tau_g$ that defines $T_g$ (typically $\tau_g \equiv 100$s) and the putative position of the characteristic time-scale $\tau_\beta(T_g)$ of the $\beta$ relaxation at $T_g$, respectively; due to the merging, the relative shift between the $\alpha$ and $\beta$ relaxation time-scales should be similar near $T_g$ for the low and high $T_g$ systems, as shown in the figure. The activation enthalpies representing the $\alpha$ and $\beta$ relaxations at $T_g$ are found by extrapolating an Arrhenius form ($\tau_{\alpha,\beta}(T_g) = \tau_0 e^{\Delta H_{\alpha,\beta}/RT_g}$) to intersect the definition of $\tau_g$ (at $T_g$) for the $\alpha$ relaxation and $\tau_\beta(T_g)$ for the $\beta$ relaxation, respectively, as shown by the thin coloured lines. Thus, $\Delta H_\alpha(T_g)$ is determined from the slope of the solid line joining $\tau_\alpha = \tau_0$ at the high-$T$ limit with $\tau_\alpha = \tau_g = 100$ s at $T = T_g$. Correspondingly, $\Delta H_\beta(T_g)$ is estimated from the slope of the solid line joining $\tau_\beta = \tau_0$ at the high-$T$ limit with $\tau_\beta(T_g)$.

The barrier ratio is given by $\mathcal{R} = \Delta H_\alpha(T_g)/\Delta H_\beta(T_g) = (\ln\tau_g/\tau_0)/(\ln\tau_\beta(T_g)/\tau_0)$, and assuming that $\tau_0$ is approximately the same for both the $\alpha$ and $\beta$ relaxations, independent of $M$, the $M$-dependence of $\mathcal{R}$ is thus given by the $M$-dependence of $\tau_\beta(T_g)$. Thus, for polymers for which the $\alpha$ and $\beta$ relaxations are merged across the $M$ range, we expect to observe a very flat $\tau_\beta(T_g)$, and thus an effectively fixed $M$-independent $\mathcal{R}$. We also note that for highly flexible polymers, the variation in $T_g$ with $M$ is relatively small, which by itself also leads to smaller variations in the barrier ratio.

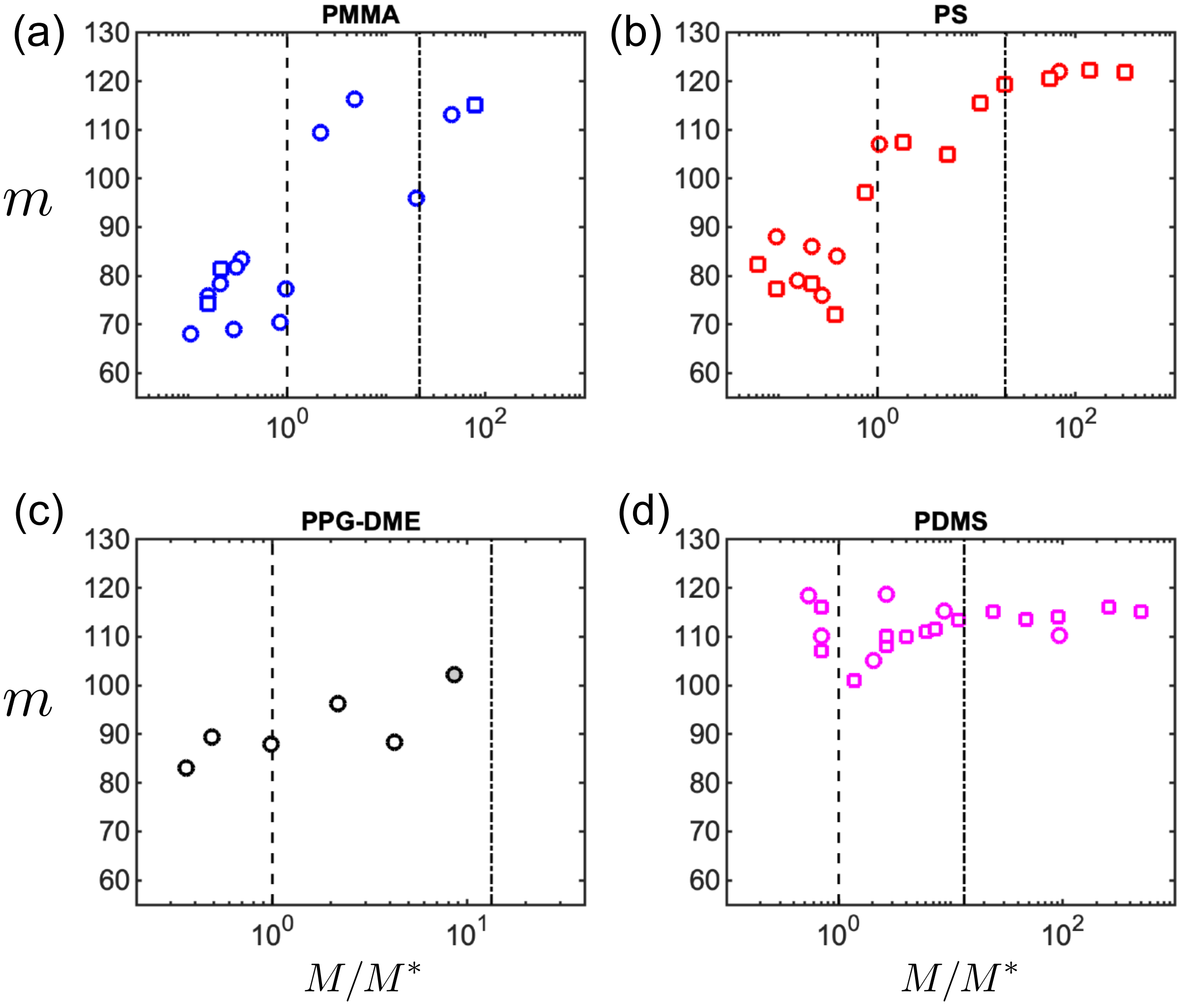

FIG. S12. The fragility $m$ for the four investigated polymers: (a) PMMA, (b) PS, (c) PPG-DME and (d) PDMS, as a function molecular weight $M$, normalised by the molecular weight $M^\star$ at which the polymer dynamics and structure turn from regime I to regime II. Our data for PMMA are shown in blue circles while data from [36] in blue squares; our data for PS are shown in red rings and data from [7] are shown in red squares; our data for PPG-DME ($n \geq 2$) are shown in black rings, while grey-filled marker shows PPG data ($M_w = 4048$ g/mol) as an indication of the long-chain behaviour [28]; our data for PDMS are shown in magenta rings, and data from [7] are shown in magenta squares.

**Fragility**— Finally, we note that it has been previously suggested that $\tau_\beta(T_g)$ is directly correlated with dynamic fragility for many glass-formers [37, 38] (as manifested in the fragility index $m = d\log(\tau_\alpha)/d(T_g/T)$), so that a more fragile glass-former (larger $m$) typically shows a smaller value of $\tau_\beta(T_g)$. Interestingly, PMMA and PS, which show clear regime behaviour in $\mathcal{R}(M)$, also show a regime behaviour in the fragility index $m$ (Fig. S12), with an essentially flat $M$-independent fragility in regime I, an increase in fragility in regime II, and a flat $M$-independent fragility in regime III [7, 36, 39, 40]. By contrast, PPG-DME shows an essentially $M$-independent fragility [28, 34] and the fragility of PDMS varies only weakly with $M$ [7, 36, 39, 40]. Thus, these observations are consistent with the lack of any significant regime behaviour observed in $\mathcal{R}(M)$ for PPG-DME and PDMS.

In conclusion, we find that the essentially flat $M$-independent behaviour of the number of correlated monomers $N_c(M)$ observed for the highly flexible PPG-DME and PDMS (see Fig. 3 in the main paper) appears also to be reflected in a similar essentially flat $M$- independent behaviour of $\mathcal{R}$, as well as in their fragility. Our results thus suggest that the observed link between the behaviour of $N_c(M)$ and $\mathcal{R}(M)$ is of a general nature.

## F. TABLE OF POLYMER DATA

| PMMA | | | | | PS | | | | | PPG-DME | | | | | PDMS | | | | |
|---|---|---|---|---|---|---|---|---|---|---|---|---|---|---|---|---|---|---|---|
| $n$ | $M_w$ | PDI | $T_\alpha$ | $T_g$ | $n$ | $M_w$ | PDI | $T_\alpha$ | $T_g$ | $n$ | $M_w$ | PDI | $T_\alpha$ | $T_g$ | $n$ | $M_w$ | PDI | $T_\alpha$ | $T_g$ |
| 2 | 202 | 1.00 | 170.3 | 166.4 | 1 | 162 | 1.00 | - | 140.1 | 1 | 104 | 1.00 | 118.0 | 113.1 | 2 | 237 | 1.00 | 122.8 | 119.2 |
| 3 | 302 | 1.00 | 214.0 | 209.8 | 2 | 266 | 1.00 | 201.0 | 197.8 | 2 | 162 | 1.00 | 141.0 | 136.1 | 3 | 311 | 1.00 | 128.2 | 124.5 |
| 4 | 402 | 1.00 | 242.3 | 238.9 | 3 | 370 | 1.00 | 235.4 | 232.3 | 3 | 220 | 1.00 | - | 149.2 | 14 | 1150 | 1.27 | 141.0 | 137.6 |
| 6 | 660 | 1.21 | 258.6 | 253.7 | 5 | 545 | 1.16 | 240.6 | 238.4 | 7 | 452 | 1.02 | 174.0 | 170.4 | 20 | 1600 | 1.37 | 141.7 | 138.7 |
| 7 | 771 | 1.18 | 274.2 | 268.5 | 7 | 725 | 1.09 | 274.2 | 273.5 | 16 | 974 | - | - | 186.8 | 66 | 4980 | 1.29 | 146.0 | 142.9 |
| 8 | 840 | 1.44 | 286.2 | 276.1 | 9 | 970 | 1.12 | 275.2 | - | 32 | 1902 | - | - | 195.3 | 623 | 46200 | 1.12 | 148.0 | 144.4 |
| 11 | 1100 | 1.17 | 299.7 | - | 18 | 1920 | 1.08 | 332.6 | 332.3 | 69 | 4048 | 1.05 | 204.0 | 200.5 | | | | | |
| 19 | 1900 | 1.10 | 333.7 | 333.7 | 28 | 2960 | 1.04 | 346.9 | - | | | | | | | | | | |
| 43 | 4300 | 1.05 | 363.8 | 365.3 | 108 | 11300 | 1.02 | 369.5 | - | | | | | | | | | | |
| 96 | 9590 | 1.05 | 373.4 | 377.0 | 456 | 47500 | 1.03 | 377.3 | - | | | | | | | | | | |
| 395 | 39500 | 1.04 | 397.7 | 390.1 | 1159 | 121k | 1.04 | - | 373.6 | | | | | | | | | | |
| 906 | 90600 | 1.04 | 394.9 | 392.7 | 30k | 3.15M | 1.05 | 379.1 | - | | | | | | | | | | |

TABLE I. Sample details, including the polydispersity index PDI. Molecular weights $M_w$ are in g/mol and temperatures $T_g$ and $T_\alpha$ are in Kelvin. The temperature $T_\alpha$ is determined using TMDSC where $\tau_\alpha = P/2\pi \approx 10$s ($P = 60$ s is the modulation period) and $T_g$ is determined from BDS VFT fits using the criterion $\tau_\alpha(T_g) = 100$s.